\documentclass[trackchanges]{aastex701}
\usepackage{amsmath}
\usepackage{longtable}

\newcommand{\kms}{km\,s$^{-1}$}
\newcommand{\ms}{m\,s$^{-1}$}

\submitjournal{ApJ}

\begin{document}

\title{Polarization Observations of a sample of Excited OH masers }

\author[0000-0003-3593-9707,gname=Derck, sname=Smits]{Derck P. Smits}
\affiliation{UNISA Centre for Astrophysics and Space Sciences (UCASS), College 
of Science, Engineering and Technology, University of South Africa, Florida 
1709, South Africa}
\email[show]{derck.smits@gmail.com}

\author[0000-0002-4846-1741,gname=Paul, sname=Fallon]{Paul Fallon}
\affiliation{UNISA Centre for Astrophysics and Space Sciences (UCASS), College 
of Science, Engineering and Technology, University of South Africa, Florida 
1709, South Africa}
\affiliation{Centre for Space Research, North-West University, Private Bag X1290, 
Potchefstroom 2520, South Africa}
\email[show]{pauljfallon13@gmail.com}

\shortauthors{Smits \& Fallon}


\begin{abstract}

Spectra of 4.7 and 6.0\,GHz excited OH masers from 21 pointings towards known 
star-forming regions are reported. The $C$-band observations, using the Green 
Bank Telescope in full Stokes mode, have measured how polarization properties 
vary across the maser profiles in each spectrum and vary between different 
epochs of observation. Seven sources had no detections of masers at any of 
the OH transitions. Nine sources were observed to have 4.766\,GHz masers, 
which include two new detections in G188.946+0.886 and G196.454--1.677.
No new linear or circular polarization was found in any of these masers. 
Seven and two sources had 6.035 and 6.031\,GHz masers, respectively. All 
6.035\,GHz masers had some level of circular polarization, and three also 
had linear polarization. Both linear and circular polarization were found 
in the 6.031\,GHz masers in G133.947+1.064, and weak, 100\% LCP in 
G141.918+1.902. Magnetic fields have been calculated from measured Zeeman 
splittings. The strength and orientation of the fields agree with most 
previous measurements, although there are some indications that the magnetic 
fields could be changing. Thermal emission or absorption is seen in several 
of the 4.7 and 6.0\,GHz transitions, many of which have not been reported before.
\end{abstract}


\section{Introduction}\label{sec:intro}

\defcitealias{SF25}{Paper~I} \defcitealias{FS26}{Paper~II}
The energy levels in the OH molecule are split into two chains of rotational 
transitions labeled $^2\Pi_{3/2}$ and $^2\Pi_{1/2}$. Lambda doubling and 
hyperfine interactions split each rotational level into four sublevels. The 
first excited level of OH is a rotational level of the $^2\Pi_{3/2}$ chain 
with $J = 5/2$. Transitions between all four sublevels, which have total 
angular momenta of $F=2$ and 3, are allowed. Masers from the $F = 3 
\rightarrow 3$ 6.035\,GHz mainline transition occur about 3 times more 
frequently than the $F = 2 \rightarrow 2$ 6.031\,GHz transition; for example, 
\citet{AQF16}, \citet{SWB20} and \citet{OCS22} detected 127, 37 and 30 masers 
at 6.035\,GHz  but only 36, 10 and 9 objects with 6.031\,GHz in their surveys. 
Only one weak 6.049\,GHz satellite line maser \citep{BDW97} has been detected 
thus far, and also only one 6.017\,GHz maser \citep{QOS25}. 

The second excited level of OH is the lowest rotational level of the 
$^2\Pi_{1/2}$ chain, which has total angular momenta $F = 0, 1$ for the lower 
(+) and upper (--) levels. Because the $F = 0 \rightarrow F = 0$ transition 
is strictly forbidden, there are only three allowed transitions in this 
rotational level. Only $\sim 50$ sources of 4.7\,GHz masers have been 
discovered in star-forming regions; see Table~\ref{tab:discover4.7} in 
Appendix A. To date, only nine 4.660\,GHz masers have been reported, and 
there are only two detections of 4.751\,GHz masers. Any new discoveries of 
these rare masers would be most welcome and could help elucidate why these 
transitions are so rare in Nature. Monitoring of these masers by \citet{S97}, 
\citet{SKH00} and \citet{S03} found that all these 4.7\,GHz sources are 
variable on time scales of months to yrs. The 4.766\,GHz maser in Mon R2 
underwent spectacular flaring \citep{SCH98} between 1994 and 1998 and has had 
a few more outbursts since then \citep{FZS06}; \citepalias{SF25}. 

The origin of these flux changes has not been established. However, some 
variation of excited OH (exOH) masers has been seen in NGC6334 I in 
association with a supposed accretion event \citep{MSG18}. It appears that 
methanol masers can be an indicator of these accretion events, but clearly 
other species of masers are also affected by the injection of energy associated 
with an accretion event. There is not enough evidence yet to prove that the 
flaring events seen in Mon R2 are due to accretion events in the protostar, 
but if they are, they have very different characteristics to those seen when 
6.7\,GHz methanol masers flare. In particular, the 4.766\,GHz maser flares 
are not accompanied by simultaneous flaring in the 6.7\,GHz methanol masers. 
The short lifetimes of many 4.7\,GHz masers might mimic the light curve of 
the exOH in NGC6334 I and exist only as transient events of several weeks. 
Monitoring of a set of these masers might be able to shed some light on 
their behavior.

Most levels of OH are paramagnetic and produce Zeeman splitting of lines in 
milliGauss strength fields. Observations of OH masers have provided important 
measurements of the magnetic field $B$ in star-forming regions by studying 
the amount by which maser lines are separated due to the Zeeman effect (see 
\citet{E92}, for example). The lowest rotational level of the $^2\Pi_{1/2}$ 
chain with $J = \frac{1}{2}$ is, unlike all other rotational levels, 
diamagnetic, and has a small Land\'{e} $g$-factor and requires fields of tens 
of mG to produce observable Zeeman shifts \citepalias{SF25}. Therefore, it 
was unexpected when \citet{SCH98} reported linear polarization in the flaring 
Mon R2 4.766\,GHz exOH masers. Circular polarization has also recently been 
detected in two 4.766\,GHz masers as reported in \citetalias{SF25}. Although 
the Land\'{e} $g$-factor is uncertain, the fields measured from the Zeeman 
shift in these circularly polarized masers align with magnetic field magnitudes 
measured from Zeeman shifts in 6.7\,GHz methanol masers reported by \citet{SVL22} 
and in \citetalias{FS26}.

Very few recent surveys have been done with a full set of Stokes parameters 
so it is difficult to know what the polarization characteristics of these 
masers are. Mon R2 is the only 4.766\,GHz maser reported to have any linear 
polarization, and now two sources have displayed circular polarization. For 
the 6.0\,GHz exOH masers there is little information on the linear polarization 
characteristics of these sources, in particular whether these characteristics 
change with time. Circular polarization can be used to calculate line-of-sight 
(LOS) magnetic fields $B_{\parallel}$ if Zeeman pairs can be identified in the 
right and left circular polarization (RCP and LCP) spectra.

This is Paper III in a series that is using data from project AGBT22B-354 that 
looked at 21 pointings towards 19 sources with the $C$-band of the 100m GBT in 
full Stokes mode. In \citet{SF25} \citepalias{SF25} we reported the first 
detection of circular polarization in 4.766\,GHz exOH masers towards two sources, 
and in \citet{FS26} \citepalias{FS26} we reported our observations of 6.7\,GHz 
methanol masers in this sample, most of which display polarization of a few 
percent in both linear and circular modes, varying across the spectra and with 
time. In this paper we present the results of observing 4.7 and 6.0\,GHz exOH 
masers towards the sources, obtaining both linear and circular polarization 
parameters. Besides the maser emission, we also confirm the presence of broad 
thermal emission and absorption lines in several of these sources, and report 
some new detections as well.

\section{Observations}
We observed 21 pointings with Right Ascensions (RA) in the range from 23$^h$ 
to 08$^h$ to meet the Local Sidereal Time range for filler sources specified 
by a GBT Special Proposal Call. Table~\ref{tab:source} lists the objects and 
their J2000 coordinates, the central velocity of the spectra, the linear 
polarization calibrator used, the time on source, and the dates when the 
observations were done. Most sources were only observed once, but where 
observing opportunities allowed, sources were observed twice to see if flux 
density, central velocity or polarization properties changed in the months 
between observations. The source \object{G213.705--12.60} (\object{Mon R2 IRS 3}) 
was observed as part of this programme initially, but when it started flaring 
at 4.766\,GHz, it was observed multiple times. The time series results of 
these data will be reported separately.

\begin{table*}
	\centering
	\caption{Source list in order of increasing RA containing source 
        name, J(2000) coordinates, velocity, polarization calibrator, 
        observing time, and the dates observed.}
	\label{tab:source}
	\begin{tabular}{clcccccc} 
		\hline
	No.	&Source   &R.A.(J2000)   &Dec.(J2000)   &Velocity   &Polarization  
    &\multicolumn{1}{c}{Time}    &Date Observed   \\
         &        &(hh mm ss.s)    &$\degr\ \ '\ \ ''$   &(\kms)  &Calibrator 
         &\multicolumn{1}{c}{(minutes)}  &(20yy mmm dd)      \\
		\hline
	  1	 &G108.758-0.986   &22 58 47.5   &+58 45 01.8   &--46   &3C48  &50  &23 Jul 31 \\
    2a &G111.526+0.803   &23 13 33.1   &+61 29 15.2   &--59   &3C48  &68  &23 Apr 21 \\ 
    2b &                 &             &              &       &      &42  &23 Jun 23 \\
    3  &G111.532+0.759   &23 13 43.9   &+61 26 55.7   &--59   &3C48  &62  &23 Jul 07 \\
    4  &G111.542+0.777   &23 13 45.3   &+61 28 10.0   &--58   &3C48  &60  &23 Aug 03 \\
    5  &G126.715--0.822  &01 23 33.7   &+61 48 49.2   &--12   &3C48  &64  &23 Nov 21 \\
    6  &G133.715+1.215   &02 25 40.6   &+62 05 50.5   &--40   &3C48  &64  &23 Dec 24 \\
    7  &G133.947+1.064   &02 27 03.7   &+61 52 25.0   &--44   &3C48  &68  &23 Jul 28 \\
    8  &G141.918+1.902   &03 27 28.6   &+58 53 48.1   &\ --8  &3C48  &64  &23 Jul 07 \\
    9  &G163.078--1.926  &04 49 46.7   &+41 39 05.5   &--15   &3C138 &64  &23 Jul 06 \\
    10 &IRAS 05137+3919  &05 17 12.8   &+39 22 05.0   &--20   &3C138 &64  &23 Dec 31 \\
    11 &G173.482+2.446   &05 39 13.0   &+35 45 51.0   &--16   &3C138 &64  &23 Apr 23 \\
    12 &G211.567--19.28  &05 39 56.0   &--07 30 27.7  &+2     &3C138 &68  &23 Apr 21 \\
    13 &IRAS 05382+3547  &05 41 37.4   &+35 48 49.0   &--24   &3C138 &68  &23 Apr 24 \\ 
    14a &IRAS 05392--0214 &05 41 44.8   &--02 13 22.7  &+11   &3C138 &68  &23 Apr 22 \\
    14b &                 &             &              &      &      &64  &23 Jun 24 \\
    15 &G183.349--0.575  &05 51 11.0   &+25 46 15.9   &\ --6  &3C138 &64  &23 Jul 06 \\
    16a &G213.705--12.60 &06 07 47.7   &--06 23 01.2  &+11    &3C138 &70  &23 Dec 23 \\
    16b &                &             &              &       &      &70  &20 Feb 24\\
    17 &G189.030+0.783   &06 08 36.1   &+21 20 28.0   &\ +3   &3C138 &60  &23 May 30 \\
    18 &G188.946+0.886   &06 08 53.3   &+21 38 29.0   &+11    &3C138 &80  &23 May 29 \\
    19 &G196.454--1.677  &06 14 37.1   &+13 49 36.0   &+15    &3C138 &60  &23 Jul 28 \\
    20 &G232.620+0.996   &07 32 09.8   &--16 58 13.0  &+23    &3C138 &64  &23 Jul 06 \\
   21a &G240.316+0.071   &07 44 53.2   &--24 07 39.0  &+63    &3C138 &80  &23 Oct 25 \\
   21b &                 &             &              &       &      &88  &24 Jan 27\\    
		\hline
	\end{tabular}
\end{table*}

All observations reported here were made with the 100m GBT using the $C$-band 
receiver with the Versatile GBT Astronomical Spectrometer (VEGAS) backend in 
full Stokes mode. The FWHM beam size is approximately 2.1 arcmin at 6.035\,GHz 
and 2.7 arcmin at 4.766\,GHz. The VEGAS backend contains eight bands each of 
which was operated in mode 15, providing a bandwidth of 11.72\,MHz over 32\,768 
channels, corresponding to a frequency resolution of 357.7\,Hz.  This gives a 
velocity resolution of 0.0225\,\kms\ at 4.766\,GHz and 0.0178\,\kms\ at 6.035\,GHz. 
In frequency-switching mode the velocity coverage is $\sim730$\,\kms\ at 
4.766\,GHz and $\sim580$\,\kms\ at 6.035\,GHz. Details of the observational 
procedures and data analysis are described in \citetalias{SF25}. 

The rest frequencies of the exOH lines used in seven of the eight bands of the 
VEGAS backend were 4.765 562 ($F = 1\rightarrow 0$), 4.750 656 ($F = 1 \rightarrow 
1$), 4.660 242\,GHz ($F = 0 \rightarrow 1$), 6.016 746 ($F = 2 \rightarrow 3$), 
6.030 747 ($F = 2 \rightarrow 2$), 6.035 092 ($F = 3 \rightarrow 3$), and 6.049 
084\,GHz ($F = 3\rightarrow 2$) for the $^2\Pi_{1/2}\ J = \frac{1}{2}$ and 
$^2\Pi_{3/2}\ J = \frac{5}{2}$ exOH lines. Henceforth, these transition frequencies 
will be referred to as 4.766, 4.751, 4.660, 6.017, 6.031, 6.035 and 6.049. The 
eighth band was used to look at the 6.7\,GHz methanol line, the results of which 
are presented in \citetalias{FS26}.

The Stokes spectra are obtained using orthogonal linear feeds and calibrated 
using a Mueller matrix as described in \citet{FSG23}. A standard linear 
polarization calibrator, either 3C48 or 3C138, was observed as part of each 
observing session to confirm and refine the calibration. The calibration 
process enables linear polarization fraction accuracy of $\sim 0.5\%$ and 
position angle accuracy of $\sim1$\degr\ for the reference calibrators with 
intensity of 3 to 6 Jy \citep{FSG23}. In addition to the Mueller matrix 
calibration, further refinements of the Stokes $V$, described in \citetalias{SF25}, 
are done to ensure that leakage of Stokes $I$ into Stokes $V$ is corrected. 
This results in an $I$ into $V$ leakage fraction calibration of better 
than $0.5\%$. 

The spectra are recorded every 2 minutes. Analysis involves shifting and 
folding each 2 minute scan, an initial intensity calibration and background 
subtraction using a third order polynomial. The Mueller matrix calibration 
is applied to each scan which are then added to produce the on-sky $I$, $Q$, 
$U$ and $V$ spectra. 

The $YX$ cross-correlation product of the spectrometer produces the Stokes $V$ 
spectrum. The gain of $V$ is not checked by our Mueller matrix calibration 
process as this uses standard linear polarization sources with $V$ assumed 
to be zero \citep{FSG23}. Some of our 6.0\,GHz spectra have non-overlapping 
RCP or LCP profiles and initially displayed regions with $V/I>100\%$ 
(and negative RCP or LCP) indicating further calibration of $V$ is required. 
It is likely that the $YX$ product has a different gain to the $XX$ and $YY$ 
auto-correlation products which are used for the Stokes $I (= XX + YY)$  
calibration. Confirming the $V$ gain calibration would require further 
observations of a source with known or calibrated $V/I$.  However, using our 
observed 6\,GHz spectra with regions where $V/I>100\%$ and testing gain 
adjustments for Stokes $V$ to have a maximum value of 100\%, results in a 
calculated gain adjustment of $0.83(1)$. This additional Stokes $V$ calibration 
has been applied to the exOH spectra in this paper. It has little impact on 
the 6.0\,GHz exOH maser magnetic field calculations when RCP and LCP profiles 
have clear separations. However, for the 4.7 and 6.7\,GHz maser magnetic fields 
determined  in \citetalias{SF25} and \citetalias{FS26}, where there is a small 
separation between the velocity peaks of the RCP and LCP profiles, this 
additional calibration means that the magnetic fields will be adjusted 
by $\sim 0.83$. Note that this assumes the $V$ gain adjustment has a limited 
frequency dependence and is similar at 4.7 and 6.7\,GHz.

\section{Results}
We have fitted multiple Gaussian profiles (using the least squares routine in 
GBTIDL) to the features in our Stokes $I$ spectra attempting to leave only noise 
as residuals. These parameters are listed for the 4.7\,GHz sources in Appendix 
\ref{ApA} Table~\ref{tab:GaussFit47} and for the 6.0\,GHz sources in Appendix 
\ref{ApB} Table~\ref{tab:GaussFit6}. Where residuals are left over that are 
larger than the noise, this is usually an indication that there are more spots 
of emission than the number of Gaussians used for the fitting. 

When there is no polarization detectable towards a source, only the Stokes $I$ 
component has been plotted. The plots use a black solid line for the original 
data and a red solid line for the fitted profiles. Rather than show the Stokes 
$Q$ and $U$ components for linear polarization, we have converted the parameters 
into percentage (\%) linear polarization ($P_\textrm{l}$), plotted in magenta, 
and position angle (P.A.), plotted in cyan. Stokes $V$ and percentage circular 
polarization ($P_\textrm{c}$) are plotted in dark green. $P_\textrm{l}$, P.A. 
and $P_\textrm{c}$ values are plotted when the Stokes $Q$, Stokes $U$ or Stokes 
$V$ spectrum is above a $3\sigma$ detection level, and the rms noise is below 
5\% of Stokes $I$. When RCP and LCP spectra are presented they are plotted in 
blue and dashed magenta, respectively. Linear and circular polarization are 
determined from the Stokes spectra in the standard manner using 
\citep{RH21} 
\begin{align}
\text{Linear polarization fraction} &= P_\textrm{l} =\frac{\sqrt{Q^2 + U^2}}{I}, 
      \text{ with} \\
\text{Position angle (P.A.)} &= \chi =\frac{1}{2} \tan^{-1}\left(\frac{U}{Q}\right), 
      \text{ and} \\
\text{Circular polarization fraction} &= P_\textrm{c} =\frac{V}{I} , \\
\text{RCP} = (I + V)/2 \quad\quad &\text{and} \quad\quad \text{LCP} = (I - V)/2 \ .
\end{align}

For cases where the Gaussian profile is a distinct peak in the maser spectrum 
and the FWHM is in the region of 0.3 \kms, it is assumed that the profile 
represents a single maser spot. If the central velocities of RCP and LCP 
profiles are equally shifted in opposite directions from the Stokes $I$ 
velocity, or there is clear separation between similarly shaped RCP and LCP 
spectra, it is likely that the separation is the result of Zeeman splitting 
$\Delta V_\textrm{Z}$. Gaussian profiles fitted to RCP and LCP are used to 
determine the peak velocities and separation.

\subsection{Non-detections} 

Of our 21 pointings, seven sources had no detections of either 4.7 or 6.0\,GHz 
exOH emission to a $3\sigma$ level of the unsmoothed rms noise ($\sim23$\,mJy 
for the 4.7\,GHz frequencies and $\sim28$\,mJy for the 6.0\,GHz frequencies). 
The sources with non-detections are: \object{G126.715-0.822}, \object{G163.078-1.926}, 
\object{IRAS 05137+3919}, \object{G211.567-19.28}, \object{IRAS 05382+3547}, 
\object{IRAS 05392-0214}, and \object{G232.620+0.996}. There have been no previous 
reports of exOH detections in G126.715-0.822, IRAS 05137+3919, IRAS 05382+3547 or 
G232.620+0.996 so we will not discuss these sources any further. The only signal in 
\object{G173.482+2.446} was from a 4.766\,GHz exOH maser that showed circular 
polarization and has been presented in detail in \citet{SF25}, so it will not be 
discussed further here.

G163.078--1.926 was observed by \citet{OCS22} to have broad lines with a width 
of several \kms\ at 6.035\,GHz. The spectrum they show in their Figure~3 shows 
a P-Cyg profile with an emission peak at $-14$\,\kms\ and an absorption dip 
at $v = -10$\,\kms. These profiles often indicate the presence of a wind from a 
stellar source. No masers have been reported towards this source. We found no 
signals above our noise level at any of the observed frequencies. 

\citet{OCS22} reported 6.035\,GHz exOH maser emission towards G211.567--19.28 
with a flux density of 0.91\,Jy at $v = 1.99$\,\kms\ in RCP and 8.2\,Jy at $v 
= 2.42$\,\kms\ in LCP. We found no exOH emission or absorption lines above the 
noise level.

No 4.766\,GHz exOH signals have been reported in IRAS 05392--0214. 6.035\,GHz 
exOH masers with 50 (RCP) and 40 (LCP)\,mJy flux densities at a velocity of 
$v = +11$\,\kms\ were found by \citet{BDW97}. We made two sets of observations 
towards this source but found nothing at any of our observed frequencies. 
Combining the two sets of observations improved the signal-to-noise ratio (SNR) 
but there was still no detectable emission.

\subsection{G108.758--0.986 (IRAS 22566+5828)} 

No 4.7\,GHz exOH masers have been found in any searches done towards this source. 
\citet{SWB20} found two peaks of RCP and LCP 6.035\,GHz masers at velocities of 
$-45.95$ and $-45.37$,\kms\ with flux densities $S_{\nu} = 2.21$ and 0.95\,Jy, 
respectively, in their observations using the Torun 32m telescope in 2019 June 
with a velocity resolution of 0.1\,\kms. Observations in 2016 Aug using the 
Tianma Radio Telescope (TMRT) with a similar velocity resolution were reported 
by \citet{OCS22} who also detected 6.035\,GHz masers but with smaller flux 
densities. Observations by \citet{KBR25} using the enhanced Multi-Element 
Remotely Linked Interferometer Network (eMERLIN) found 12 spots of which three 
are more than 80\% circularly polarized and two are less than 50\%. One spot 
had linear polarization of 12\%. They identified two sets of Zeeman pairs which 
produced $B_{\parallel}$ towards the observer of --8.5 and --10.6\,mG.

\begin{figure}[ht!]
\centering
\scalebox{0.6}{\plotone{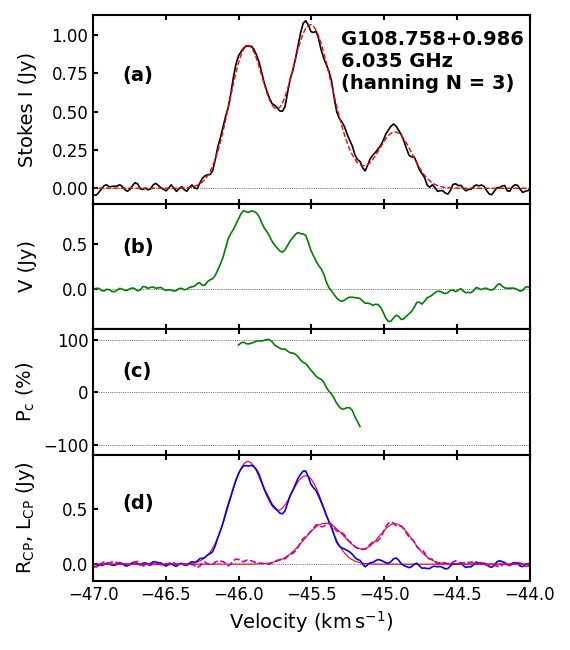}}
         \caption{Spectra of G108.758--0.986 showing (a) 6.035\,GHz Stokes $I$ and 
         the sum of the Gaussian fits to RCP and LCP in dashed red, (b) Stokes $V$, 
         (c) $P_\textrm{c}$, and (d) RCP and LCP spectra each fitted with two 
         Gaussians shown in red.}
\label{fig:G108_5}
\end{figure}

In keeping with previous searches, no 4.7\,GHz signals were found in our 
observations, and at 6.0\,GHz we only detected 6.035\,GHz exOH masers. Our 6.035\,GHz 
spectra are very similar to those measured by \citet{SWB20}, but we have a higher 
velocity resolution. The Stokes $I$ and $V$ spectra, $P_\textrm{c}$, and the RCP 
and LCP profiles are shown in Figure~\ref{fig:G108_5}. The 6.035\,GHz maser lines 
are 100\% circularly polarized. The Gaussian parameters for the RCP and LCP 
components are listed in Table \ref{tab:GaussFit6} and are the same values used to 
fit the $I$ component. The velocities measured by \citet{SWB20} are a bit different 
to ours, which could indicate changing magnetic fields threading the masering regions.  
More regular monitoring of this source in full Stokes mode would confirm if this is 
happening.

\subsection{NGC 7538 Complex (G111.526+0.803, G111.532+0.759 and G111.542+0.777)}
\label{sec:7538}

No 4.751 or 4.660\,GHz masers have been reported towards this source yet. However, 
using the Effelsberg 100m radio telescope, \citet{GMP83} discovered three 4.766\,GHz 
exOH maser lines and broad thermal emission underlying the masers. Observations with 
the Very Large Array (VLA) made by \citet{PGW84} found three clusters of maser spots 
corresponding to the velocities of the maser lines found by \citet{GMP83}. Nothing 
was found to a limit of 0.26\,Jy by \citet{CMW91}, however, in 1991 March \citet{CMC95} 
detected a 4.766\,GHz maser at $v = -59.2$\,\kms, which is at the same velocity as 
one of the masers found by \citet{GMP83}. Using MERLIN, \citet{HSC05} observed IRS 1, 
IRS 9 and IRS 11 from 1995 February to June, detecting 4.766\,GHz exOH masers only in 
IRS 1. They found maser spots at velocities of $v = -57.7, -58.74$ and $-59.03$\,\kms. 
Between 1998 June and December \citet{SKH00} made four observations towards 
G111.542+0.777, in which they detected a feature at $v = -57.2$\,\kms, and one 
tentative detection at $v =-61.5$\,\kms. A single feature was found by \citet{OCS22} 
in G111.526+0.803 with a velocity of $v = -58.12$ and slightly different flux 
densities in the RCP and LCP components in 2016 July using the TMRT. \citet{QSB22} 
identified three features towards G111.542+0.777 in 2018 July using the TMRT at 
velocities of $v = -60.41, -57.98$ and --56.97\,\kms. It is clear from this set of
observations that the 4.766\,GHz exOH masers in the NGC 7538 complex can change on 
scales of months and hence it can be classified as a variable source.

The first detection of 6.035\,GHz exOH in the NGC 7538 complex was made by \citet{RZP75} 
who found a single line at $v = -59.7$\,\kms\ towards IRS 1. No 6.031\,GHz signals were 
found. \citet{GBW84} identified two features in their 6.035\,GHz spectra at velocities 
of $v = -61.5$ and --59.5\,\kms. Two features were also detected by \citet{BDW97} at 
$v = - 60.6$ and --59.1\,\kms. Weak (20 -- 40\,mJy) emission at 6.049\,GHz with a 
velocity $v = -59.$\,\kms\ was found. \citet{FRM06} found three 6.035\,GHz maser lines 
at $v = -59.34, -58.83$ and --56.88\,\kms, the last of which displayed Zeeman splitting 
giving rise to a field of $B_\parallel = +0.9$\,mG. They also found broad thermal 
emission at 6.035 and 6.031\,GHz with flux densities $S_\nu = 0.07$ and 0.10\,Jy, at 
velocities $v = -58.7$ and --59.1\,\kms, and line widths $\Delta v = 3.1$ and 1.1\,\kms, 
respectively. Only one 6.035\,GHz maser line was found by \citet{SWB20} at $v = 
-59.5$\,\kms\ towards IRS 1, with a Zeeman splitting due to a magnetic field $B_{\parallel} 
= 0.6$\,mG. Towards G111.526+0.803 (IRS 4) a single 6.031\,GHz maser line was reported 
by \citet{OCS22} with a Zeeman splitting giving rise to a magnetic field $B_{\parallel} 
= -2.8$\,mG at $v = -58.85$\,\kms. They also found two maser lines towards G111.532+0.759 
(IRS 11) at 6.035\,GHz with velocities $v = -60.3$ and --59.5\,\kms, the latter of which 
had a Zeeman splitting due to a field $B_{\parallel} = +1.3$\,mG. Observations made by 
\citet{QOS25} towards G111.542+0.777 found two closely spaced maser lines (see their 
high resolution image) at 6.035\,GHz with the peak at $v = -59.36$\,\kms\ giving rise 
to Zeeman splitting in a field $B_{\parallel} = -0.3$\,mG.

Our observations pointed at three positions in this complex. According to 
the map of \cite{OTN04}, our pointings are centered on IRS 4 = 
\object{G111.526+0.803}, IRS 11 = \object{G111.532+0.759} and IRS 1 -- 3 = 
\object{G111.542+0.777}. Because these pointings are so close together, the 
same masers are seen in all three pointings. However, the flux density 
differs because the sources appear in different parts of the beam, and 
possibly because of flux density variations. IRS 1 is the brightest object 
in the cluster and is identified as an ultra-compact HII region. IRS 11 is 
the site of a 6.7\,GHz methanol maser discovered by \citet{PMM06}, and is 
suspected to be a younger star-forming region than that in IRS 1.

We found 4.766 and 6.035\,GHz exOH masers towards all three pointings, but 
with flux density variations, the strongest of which was towards G111.542+0.777. 
This suggests that the emission is in the IRS 1 -- 3 region.

\begin{figure}[ht!]
\centering
\scalebox{0.6}{\plotone{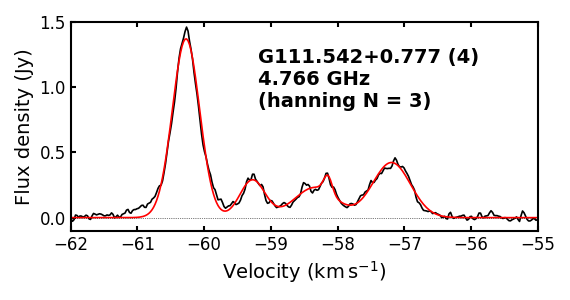}}
        \caption{Stokes $I$ spectrum of 4.766\,GHz exOH in NGC 7538 towards 
        G111.543+0.777.}
\label{fig:G111_4765}
\end{figure}

Four peaks of 4.766\,GHz maser emission were identified in all pointings, 
the profiles of which were the same at all times. There was no clear 
polarization in any of these profiles, spectra of which are shown in 
Figure~\ref{fig:G111_4765}. Parameters of the fitted Gaussians are 
listed in Table \ref{tab:GaussFit47}, together with parameters fitted to 
the other pointings in this source. Most of these profiles are broad, 
($\Delta v > 0.3$\,\kms) suggesting that these lines could be multiple 
spots of maser emission. The amplitudes and velocities of our lines are 
noticeably different from those identified by \citet{QSB22}, confirming 
the variability of the 4.766\,GHz exOH masers in this source.

\begin{figure*}[ht!] 
\plottwo{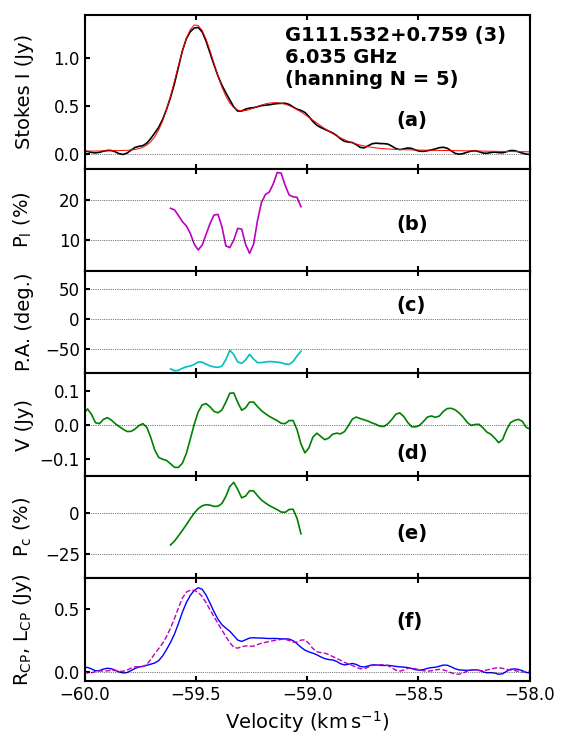}{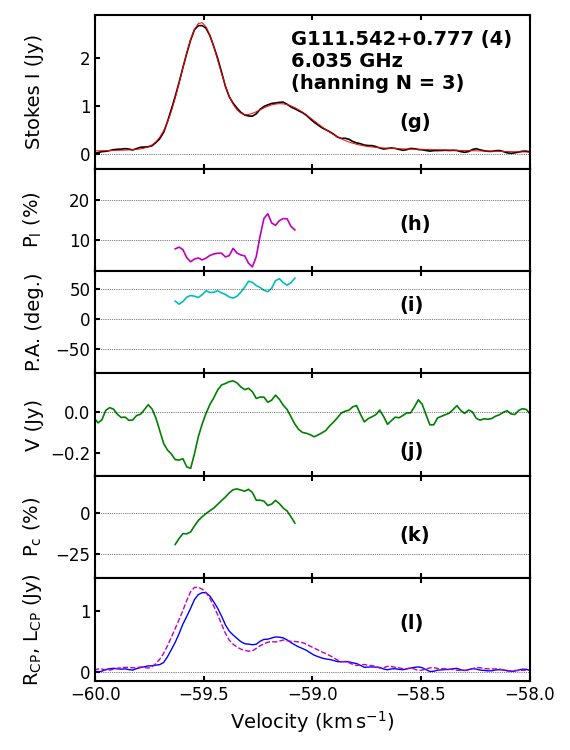}
          \caption{Spectra of the 6.035\,GHz exOH in G111.532+0.759 and 
          G111.542+0.777 showing (a) \& (g) Stokes $I$ with fitted Gaussians, 
          (b) \& (h) $P_\textrm{l}$, (c) \& (i) P.A., (d) \& (j) Stokes $V$, 
          (e) \& (k) $P_\textrm{c}$, and (f) \& (l) RCP and LCP. }
\label{fig:G111_6035B}
\end{figure*}

The 6.035\,GHz spectra for G111.532+0.759 and G111.542+0.777, shown in 
Figure \ref{fig:G111_6035B}, were resolved into two maser lines and one 
thermal emission component. As can be seen from the plots, $P_\textrm{l}$ 
and P.A. change noticeably over the month between the observations. These 
parameters should not be affected by the orientation of the antenna and, 
therefore, represent changes in the linear polarization. The circular 
polarization appears more stable, but there are small changes that are 
discussed in section \ref{sec:mag}. 

\begin{figure}[ht!]
\centering
\scalebox{0.6}{\plotone{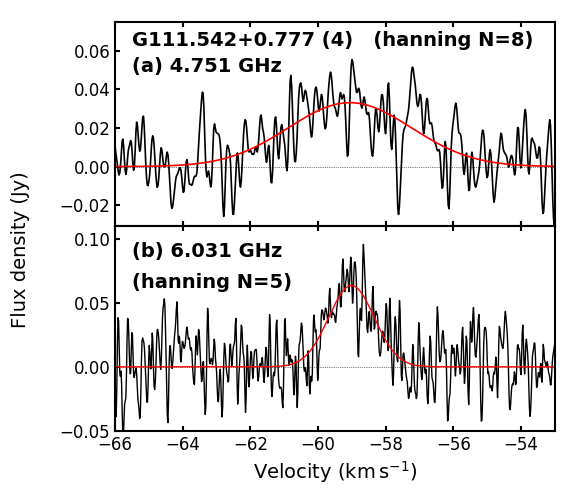}}
         \caption{Thermal emission spectra of (a) 4.751 and (b) 6.031\,GHz 
         exOH in G111.542+0.777.}
\label{fig:G111_4750}
\end{figure}

Thermal emission was also found towards G111.542+0.777 at 4.751\,GHz but not 
at 4.766 or 4.660\,GHz, as well as at 6.031\,GHz. The 4.751\,GHz emission is 
a new detection, with a single Gaussian fitted profile having the parameters 
listed in Table \ref{tab:GaussFit47}. Parameters for the 6.031\,GHz line are 
presented in Table \ref{tab:GaussFit6}. Stokes $I$ spectra of the 4.751 and 
6.031\,GHz emission are shown in Figure~\ref{fig:G111_4750}. The thermal 
components are too weak to be identified in the G111.526+0.803 spectra, 
indicating that they probably also originate in the vicinity of IRS 1 -- 3. 
The central velocity of these three features is $v = -59.1$\,kms, but the 
4.751\,GHz profile is much broader than the 6.0\,GHz components. Overall, 
this is a complicated source that warrants detailed investigation. 

\subsection{G133.715+1.215 (W3 IRS 5, IRAS 02219+6152)}  

\citet{ZP70} did not detect any 4.7\,GHz signals towards this source but when
\citet{B74} looked a few years later, a 4.766\,GHz line was found at $v = 
-34.0$\,\kms\ with a width of $\Delta v = 0.47$\,\kms. Around the same time 
\citet{RZP75} also detected emission at $v = -33.9$ and $-37.3$\,\kms. This 
latter component had a flux of $\sim 0.5$\,Jy. A decade later, \citet{GMP83} 
detected a 2.2\,Jy maser at $v = -37.4$\,\kms\ with a width $\Delta v = 
0.47$\,\kms, while the $v = -33.9$\,\kms\ component had disappeared. VLA 
observations in 1982 September made by \citet{GWP83} showed a 2\,Jy maser at 
$v = -37.4$\,\kms\ with no associated radio continuum emission. In 1989 June 
\citet{CMW91} did not detect any signals with the 76m Lovell telescope, but 
in 1991 August \citet{CMC95} identified a 1.65\,Jy maser at $v = -38.4$\,\kms\ 
using the 64m Parkes telescope. Observations made with the MERLIN array in 1995 
June by \citet{HSC05} found a 0.610\,Jy spot of maser emission at $v = -38.95$\,
\kms. The position of this maser differed from that found by \citet{GWP83} by 
$0.3''$, and the velocities of the spots are different. 

\citet{RZP75} reported broad absorption lines at 6.035 and 6.031\,GHz with a 
central velocity of $v = -40$\,\kms. Using the Effelsberg 100m telescope, 
\citet{BDW97} confirmed the existence of broad 6.0\,GHz mainline absorption. 
They found $S_{6.031} = -0.41$ and $S_{6.035} = -0.38$\,Jy at velocities of 
$v = -39.45$ and --39.71\,\kms\ with FWHM of $\Delta v = 1.33$ and 2.98\,\kms.
They also reported weak absorption at 6.017\,GHz with $S_{6.017}\sim 
-0.060$\,mJy at $v \sim -40$ and $\Delta v \sim 2.5$\,\kms.

\begin{figure}[ht!]
\centering
\scalebox{0.6}{\plotone{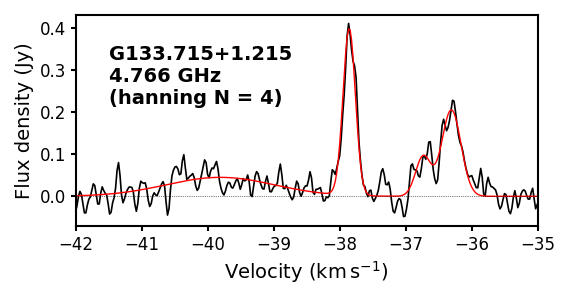}}
          \caption{Stokes $I$ spectrum of the 4.766\,GHz exOH in G133.715+1.215. }
\label{fig:G133_715_0}
\end{figure}

We found unpolarized 4.766\,GHz masers and a thermal emission feature which are 
shown in Figure~\ref{fig:G133_715_0}, and fitted with the Gaussian parameters in 
Table \ref{tab:GaussFit47}. The strongest maser line has a velocity which is 
different from previously reported values. Two other weaker features are noted, 
also at velocities not previously reported. Because these velocities do not match 
any previous identifications they must be newly developed features. The 
4.766\,GHz masers in this source appear to show high variability. The thermal 
emission centered on $v = -39.8$\,\kms\ has not been reported previously. 

\begin{figure}[ht!]
\centering
\scalebox{0.6}{\plotone{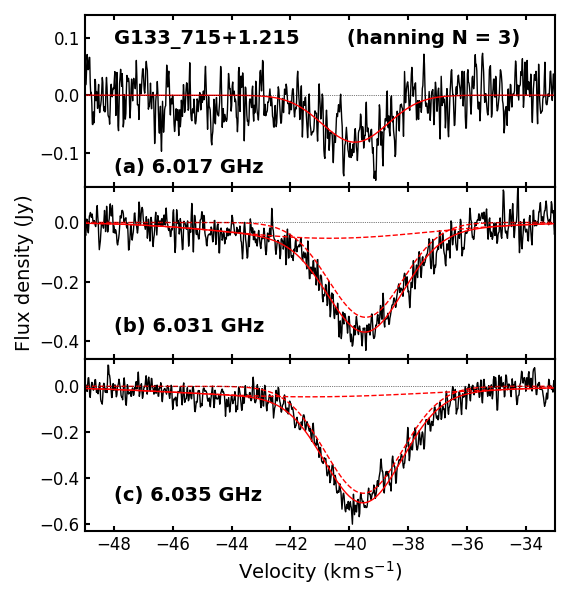}}
         \caption{Absorption spectra of (a) 6.017, (b) 6.031 and (c) 6.035\,GHz 
         in G133.715+1.215. }
\label{fig:G133_715_Abs}
\end{figure}

Absorption lines were found at 6.017, 6.031 and 6.035\,GHz, as shown in 
Figure~\ref{fig:G133_715_Abs}. Gaussian fits to these features are listed 
in Table~\ref{tab:GaussFit6}. Three Gaussians fitted to the 6.031 and 
6.035\,GHz profiles gave better overall fits than the two Gaussian fits 
shown in Figures~\ref{fig:G133_715_Abs} (b) and (c) but the uncertainties 
are so large that the amplitudes are not within 3$\sigma$ limits. The 
6.017\,GHz absorption feature has been fitted with a single broad profile 
that has the same central velocity as the 4.766\,GHz thermal emission 
feature, but is broader than this line. It has similar properties to the 
line reported by \citet{BDW97}, showing that it has been stable over 30\,yrs. 
The dips in the 6.031 and 6.035\,GHz spectra are asymmetrical, suggesting 
the presence of multiple absorption features. Each spectrum has been resolved 
into two broad components at overlapping velocities, one component of which 
is close in velocity to the central velocity of the broad lines in the 4.766 
and 6.017\,GHz spectra. These mainline absorption dips have similar 
characteristics to those identified by \citet{BDW97}, and have therefore also 
been stable over 30\,yrs. The two broad lines with FWHM of $\Delta v = 7.9$ 
and 10.1\,\kms\ in our spectra have not been reported previously. The thermal 
emission and absorption appear to arise from at least two separate clouds. 
This is one of only two sources in our sample that has 6.0\,GHz features but 
no methanol \citepalias{FS26}. 

\begin{figure}[ht!]
\centering
\scalebox{0.6}{\plotone{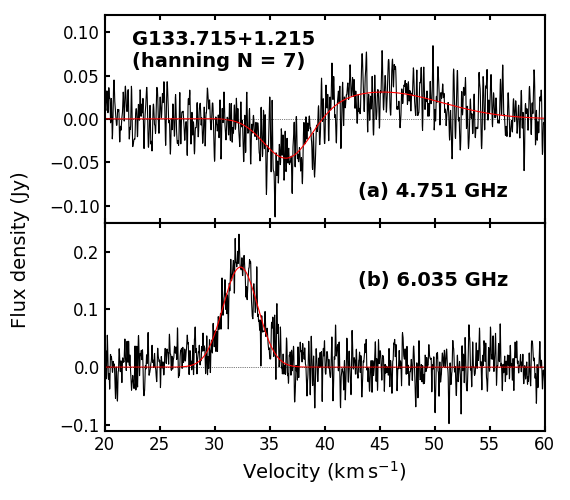}}
         \caption{The P Cyg profile at (a) 4.751\,GHz and (b) emission at 
         6.035\,GHz found in our spectra of G133.715+1.215 .}
\label{fig:G133_715_37kms}
\end{figure}

Our spectra turned up two other unusual signals that do not appear to originate 
in G133.715+1.215 because of their very different standard-of-rest velocities 
$v_\textrm{lsr}$. A P Cyg profile found at 4.751\,GHz with a dip at $v = 
+36.7$\,\kms\ and emission centered on $v = +45.2$\,\kms, and a broad emission 
profile at 6.035\,GHz at $v = +32$\,\kms\ are shown in Figure~\ref{fig:G133_715_37kms}. 
Gaussians fitted to these features are listed in Tables \ref{tab:GaussFit47} and 
\ref{tab:GaussFit6} as separate items to distinguish them from features at the 
standard $v_\textrm{lsr}$. No other signals at 4.7 or 6.0\,GHz at these velocities 
were found in our data. These must be from foreground or background sources along 
the LOS but the precise origin of these signals remains a mystery for the time being.

\subsection{G133.947+1.064 (W3(OH))} 

W3(OH) is a source rich in masers that has been observed extensively with 
both single dish telescopes and interferometers at a range of frequencies. It 
was one of the first 4.766\,GHz exOH masers discovered by \citet{ZPP68} and 
6.031 and 6.035\,GHz masers discovered by \citet{YZP69} with velocities $v 
\sim -44$\,\kms. The 4.766\,GHz masers have been fairly stable since their 
discovery, displaying a pair of maser peaks at velocities $v = -43.4$ and 
--45.0\,\kms. The most recent report is by \citet{QSB22} who fitted four 
spectral components at velocities $v = -45.1, -44.4 -43.5,$ and --43.2\,\kms. 
They reported no 4.660 or 4.751\,GHz emission.

Mainline 6.0\,GHz masers were discovered in this source by \citet{YZP69} who 
also noted circular polarization in the profiles. Follow up observations by 
\citet{ZYGP72} pointed out activity over the velocity range $v = [-49, 
-42]$\,\kms\ and the RCP and LCP components had different amplitudes. The first 
Very Long Baseline Interferometry (VLBI) experiment was carried out by 
\citet{MRL78} who identified 10 Zeeman pairs of maser spots giving rise to 
magnetic fields in the range $B_{\parallel} = [2, 9]$\,mG. Higher resolution 
European VLBI Network observations carried out by \citet{BDW97} found 61 masers 
at 6.035\,GHz and 21 at 6.031\,GHz. Of these there were 56 and 9 Zeeman pairs 
in the 6.035 and 6.031\,GHz spectra, produced by magnetic fields with 
$B_{\parallel} = 2 - 10$\,mG. They noted some variability in the spectra since 
the observations of \citet{MRL78}. The MERLIN observations of \citet{ECG05} 
found fewer spots of emission due to their lower resolution, but also noted 
some variations in the spectra. They found 12 and 5 Zeeman pairs at 6.035 and 
6.031\,GHz, respectively. The single dish spectra of \cite{SWB20} also noted 
variability compared to previous studies. The single dish observations of 
\citet{QOS25} did not identify any Zeeman pairs.

Weak, broad 6.049\,GHz emission was discovered by \citet{GBW84} with $S_{6.049} 
= 180$\,mJy, $v = -45.1(4)$ and $\Delta v = 2.1(2)$\,\kms\ and confirmed by 
\citet{BDW97} with a similar flux density and velocity, but much narrower 
line width ($S_{6.049} = 18(4)$\,mJy, $v = -44.54(2)$ and $\Delta v = 
0.86(7)$\,\kms) and pointed out an asymmetry in the line profile. They also 
noted weak, broad absorption at 6.017\,GHz with $S_{6.017} = -54(42)$\,mJy, 
$v = -45.02(10)$ and $\Delta v = 1.8(3)$\,\kms. \citet{FSP07} confirmed the
emission and absorption at 6.049 and 6.017\,GHz

\begin{figure}[ht!] 
\centering
\scalebox{0.6}{\plotone{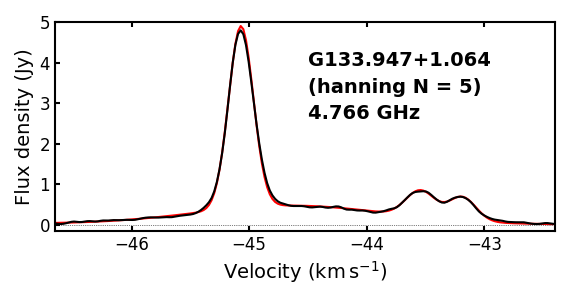}}
        \caption{Stokes $I$ spectrum of 4.766\,GHz exOH in G133.947+1.064. }
\label{fig:G133_4765}
\end{figure}

Our observations show a complex set of spectra at several different frequencies. 
Our 4.766\,GHz spectrum, plotted in Figure~\ref{fig:G133_4765}, is fitted with 
five Gaussians listed in Table \ref{tab:GaussFit47}. Three of these profiles 
are narrow enough to be masers with the same velocities found by \citet{QSB22},
while the other two profiles are broad and indicative of thermal emission. The 
maser features at $v \sim -43.4$\,\kms\ are much weaker relative to the $v = 
-45.1$\,\kms\ line than they were for the first 30 yrs after discovery \citep{SKH00}. 
None of the masers are polarized. 

\begin{figure}[ht!] 
\centering
\scalebox{0.6}{\plotone{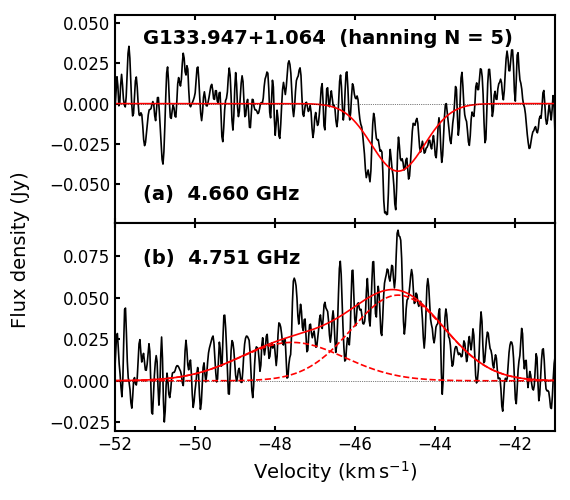}}
    \caption{Stokes $I$ spectra of the 4.660 and 4.751\,GHz exOH in 
    G133.947+1.064. Red dashed lines are the individual Gaussian profiles.}
\label{fig:G133_4660}
\end{figure}

At 4.660\,GHz an absorption dip was found, which is plotted in Figure~\ref{fig:G133_4660}. 
Also in this figure is broad band emission at 4.751\,GHz which we fitted with the two 
Gaussians listed in Table \ref{tab:GaussFit47}. The central velocity of the 4.660\,GHz 
line is the same as the stronger emission line in the 4.751\,GHz spectrum, which suggests 
that they could be formed in the same cloud. The two 4.751\,GHz emission lines have the 
same velocities within the uncertainties as features in the 4.766\,GHz spectrum. 

\begin{figure*}[ht!] 
\plottwo{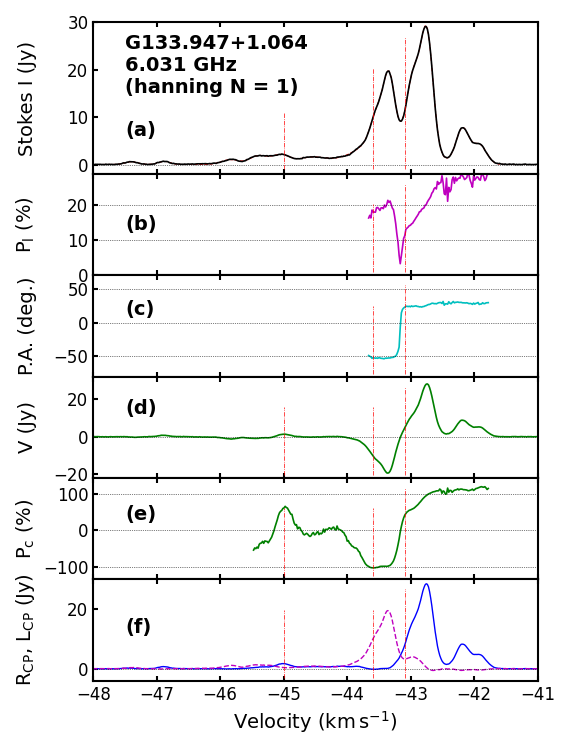}{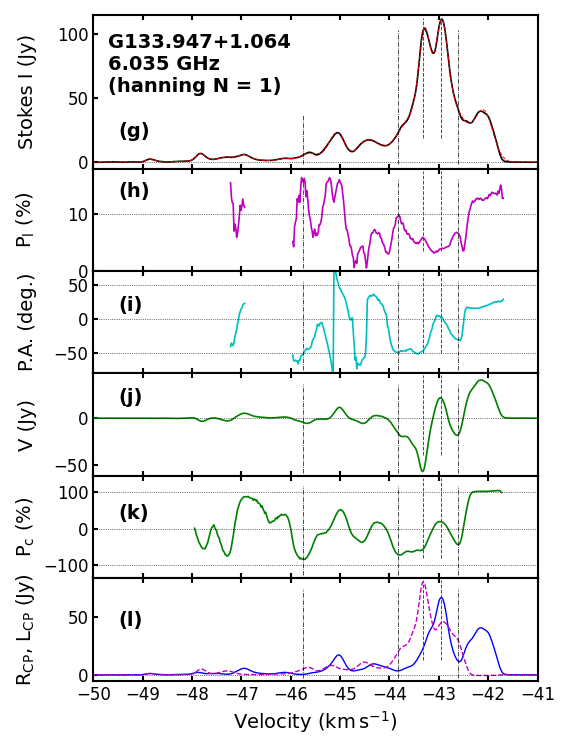}
        \caption{Spectra of the 6.031 and 6.035\,GHz exOH in G133.947+1.064 showing 
        (a) \& (g) Stokes $I$, (b) \& (h) $P_\textrm{l}$, (c) \& (i) P.A., (d) \& 
        (j) Stokes $V$, (e) \& (k) $P_\textrm{c}$, and (f) \& (l) RCP and LCP 
        profiles.}. 
\label{fig:G133_6M}
\end{figure*}

At 6.031 and 6.035\,GHz the spectra are complex, consisting of numerous lines 
overlapping in velocity with high percentages of circular polarization. Spectra 
are shown in Figure~\ref{fig:G133_6M} and parameters of Gaussians fitted to these 
are listed in Table \ref{tab:GaussFit6}. Suggested Zeeman pairs together with 
estimates of the LOS magnetic field are discussed in section~\ref{sec:mag}.

\begin{figure*}[ht!] 
\centering
\scalebox{0.6}{\plotone{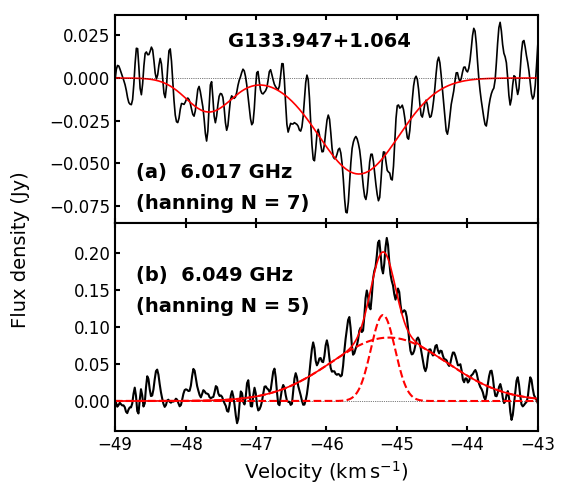} }
    \caption{Spectra of the Stokes $I$ (a) 6.017 and (b) 6.049\,GHz exOH 
    satellite lines in G133.947+1.064. Red dashed lines in (b) are the 
    individual Gaussian profiles. }
\label{fig:G133_6S}
\end{figure*}

The 6.017\,GHz satellite line has an absorption dip which was resolved into 
two broad lines that are listed in Table \ref{tab:GaussFit6}. The velocities of 
these lines do not match any of the identified 6.0\,GHz profiles in this source. 
For the other satellite line at 6.049\,GHz emission is found. Two Gaussians have 
been fitted to the emission (see Table \ref{tab:GaussFit6}), showing one 
broad profile and one with a width of $\Delta v = 0.40$\,\kms. This narrow 
width identifies this as a probable maser feature, as suggested by \citet{FSP07}. 
Spectra of these lines are shown in Figure~\ref{fig:G133_6S}. Our observations 
have much better velocity resolution than previous studies, and provide the 
first reliable values for the Gaussian parameters of the 6.049\,GHz maser. 
This maser appears to have been around for over 40 yrs, which is surprising 
given that it is the only known maser from this transition, whose rarity 
suggests unusual conditions for its existence.

\subsection{G141.918+1.902} 
Also known as AFGL490 27, this source is classified as a Young Stellar 
Object in SIMBAD, based on the {\it Spitzer} survey of this region by 
\citet{GMM09}. No 4.7\,GHz exOH masers have been reported in this source.
\citet{OCS22} found 6.031 and 6.035\,GHz exOH masers at velocities of 
$v = -8.0 \mbox{ and} -8.14$\,\kms, respectively. 

\begin{figure*}[ht!]
\plottwo{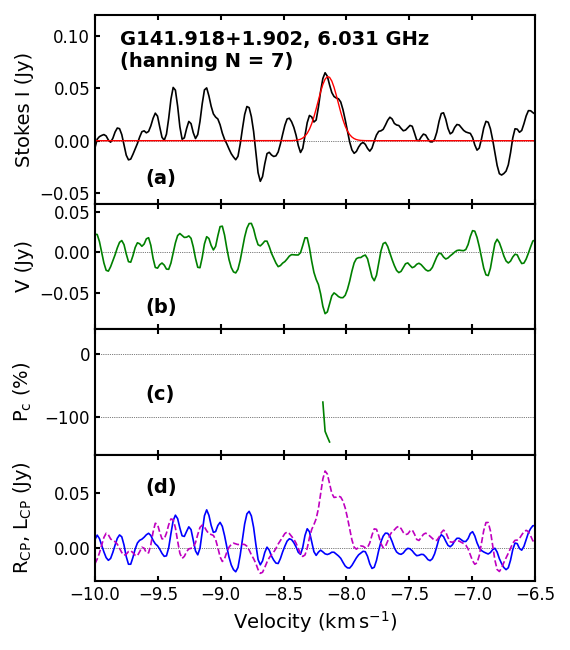}{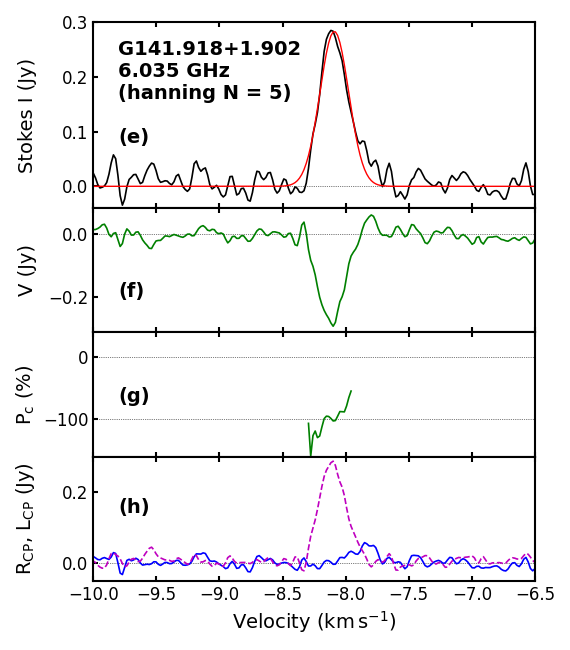}
    \caption{Spectra of the 6.031 and 6.035\,GHz exOH maser profiles in 
    G141.918+1.902 showing (a) \& (e) Stokes $I$ and fitted Gaussians, 
    (b) \& (f) Stokes $V$, (c) \& (g) $P_\textrm{c}$,, and (d) \& (h) 
    RCP and LCP profiles.}
\label{fig:G141_6031}
\end{figure*}

We found both 6.031 and 6.035\,GHz masers, as can be seen in the spectra 
presented in Figure~\ref{fig:G141_6031}. There is no linear polarization 
above our noise limits in the 6.031 or 6.035\,GHz profiles. However, at 
6.031\,GHz there is a $3\sigma$ circular polarization detection indicating 
100\% LCP with no RCP. The 6.035\,GHz spectrum, which is about six times 
stronger, is also predominantly 100\% LCP but there is a small region of 
RCP between $v = -8.0$ and --7.8\,\kms. These features appear to have 
changed since they were observed by \citet{OCS22}, who measured a magnetic 
field $B_{\parallel} = -2.8$\,mG from Zeeman splitting in the 6.031\,GHz 
lines. Measured Zeeman splitting and the calculated magnetic fields are 
discussed in section~\ref{sec:mag}. This source, like G133.715+1.215 in 
our sample of 19 objects, has 6.0\,GHz masers but no 6.7\,GHz methanol 
masers \citepalias{FS26}. It is also the only source other than 
G133.947+1.064 in our sample that has a 6.031\,GHz maser, albeit a very 
weak line.

\subsection{G183.349--0.575 (IRAS 05480+2545)}  

No 4.7\,GHz masers have been reported in this source previously, but 6.031 
and 6.035\,GHz masers were found by \citet{SWB20}. The 6.031\,GHz maser 
consisted of a single RCP line at $v = -5.15$\,\kms. Observations made in 
2016 Aug by \citet{OCS22} also found a 6.035\,GHz maser, but, as discussed 
by them, at different velocities and fluxes to those reported by \citet{SWB20}. 

\begin{figure*}[ht!]
\centering
\scalebox{0.6}{\plotone{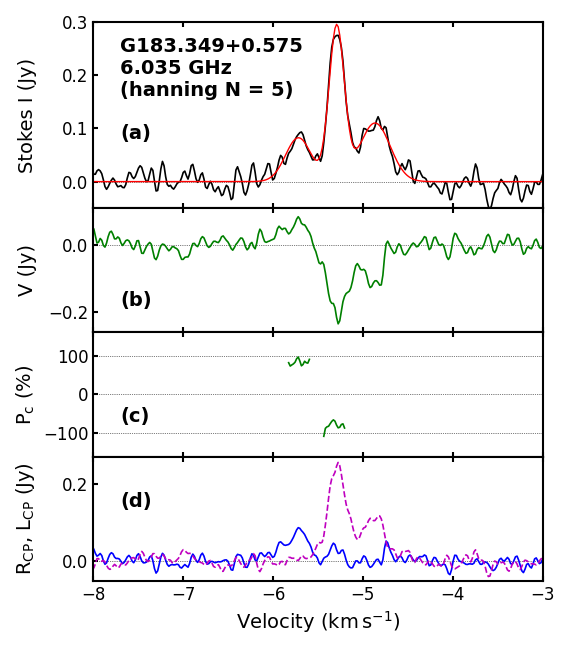}}
         \caption{Spectra of 6.035\,GHz exOH in G183.349--0.575 showing 
         (a) Stokes $I$ and fitted Gaussians, (b) Stokes $V$, (c) $P_\textrm{c}$, 
         and (d) RCP \& LCP profiles. }
\label{fig:G183_5}
\end{figure*}

We found 6.035\,GHz masers in this source, as can be seen from the spectra 
shown in Figure~\ref{fig:G183_5}. The $I$ spectrum can be fitted well with 
the three Gaussians listed in Table~\ref{tab:GaussFit6}. The spectra are 100\% 
circularly polarized and resolve into two RCP and two LCP profiles which are 
also listed in Table~\ref{tab:GaussFit6}.. The velocities of the lines we 
detected are different to those found by \citet{SWB20} and \citet{OCS22}, 
which suggests that this is a rapidly changing source of 6.035\,GHz emission. 
We have assumed that the two RCP and LCP profiles are Zeeman pairs, and in 
section~\ref{sec:mag} have calculated magnetic fields for these spots. 
Interferometric observations are needed to confirm that these are Zeeman 
pairs. We did not detect any 6.031\,GHz lines to our noise limit of 18\,mJy.  

\subsection{G213.705--12.60 (hereafter Mon R2)} 

Mon R2 underwent a major flaring episode at 4.766\,GHz in the 1990s 
\citep{SCH98, S03}. An unexpected discovery from these events was that 
the masers had 14\% linear polarization. It flared again in 2006 
\citep{FZS06}, and monitoring at the Hartebeesthoek Radio Astronomy 
Observatory since 2012 shows that it has undergone a number of flaring 
episodes. The most recent flaring episodes (see \citetalias{SF25}) will 
be reported separately (P. Fallon and D.P. Smits, in preparation). The 
first detection of circular polarization in 4.766\,GHz exOH masers has 
been reported recently in \citetalias{SF25} in this source, and hence 
will not be discussed further here. 

No 6.0\,GHz masers have been found towards Mon R2, even when the 4.766\,GHz 
exOH maser is flaring. However, observing with the Effelsberg 100m telescope, 
\citet{GBW84} reported broad absorption lines at 6.031 and 6.035\,GHz having 
a depth of $\sim -0.2$\,Jy, with the 6.031\,GHz line being slightly deeper 
than the 6.035\,GHz dip. Clearly, the pumping mechanism for the 4.766\,GHz 
transition is different to that for the 6.0\,GHz lines.

\begin{figure*}[ht!]
\centering
    \scalebox{0.6}{\plotone{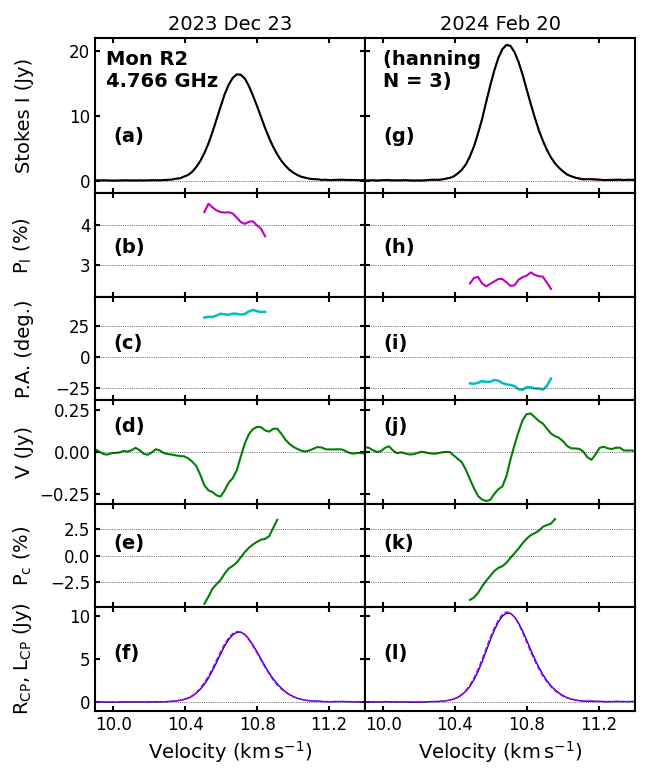}}
    \caption{Spectra of 4.766\,GHz towards Mon R2 at two epochs showing (a) 
    \& (g) Stokes $I$, (b) \& (h) $P_\textrm{l}$, (c) \& (i) P.A., (d) \& 
    (j) Stokes $V$, (e) \& (k) $P_\textrm{c}$, and (f) \& (l) RCP and LCP 
    profiles.}
    \label{fig:G213_4765}
\end{figure*}

In \citetalias{FS26}, plots of the 6.7\,GHz methanol masers on 2023 December 
23 and 2024 February 20 are presented to illustrate the changes occurring 
in $P_\textrm{l}$, P.A., and the direction of the Stokes $V$ component in 
the prominent peaks in the spectra. On those same two dates, the linear 
polarization in the 4.766\,GHz exOH maser also underwent significant changes, 
as can be seen in the plots presented in Figure~\ref{fig:G213_4765}. The 
percentage linear polarization decreases from 4.1\% to 2.6\% while the P.A. 
changes from +30\degr\ to --25\degr. 

\begin{figure*}[ht!]
\centering
    \plottwo{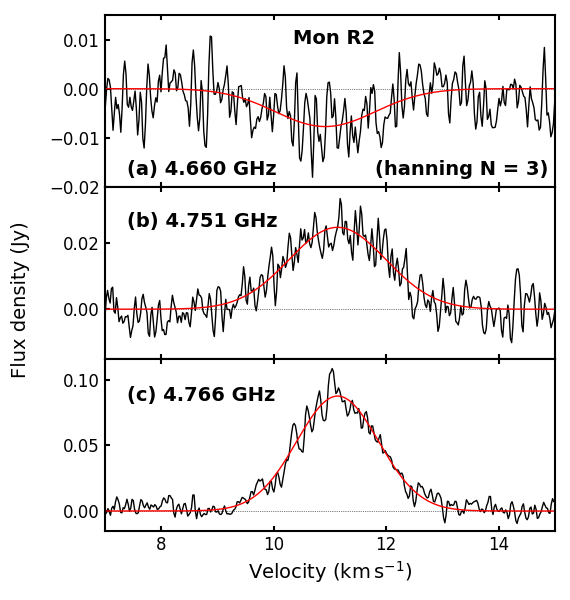} {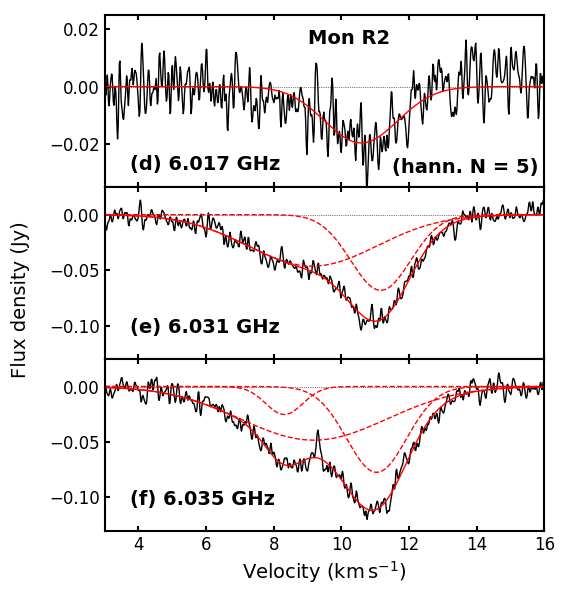} 
    \caption{Stokes $I$ spectra of Mon R2 showing (a) absorption at 4.660\,GHz 
    and emission at (b) 4.751, (c) 4.766\,GHz, and absorption lines at (d) 
    6.017, (e) 6.031 and (f) 6.035\,GHz.} 
    \label{fig:G213_67}
\end{figure*}

We report the first detection of thermal emission at 4.751\,GHz and absorption 
at 4.660, and 6.017, and confirm absorption at 6.031 and 6.035\,GHz. Because 
this source started flaring at the end of 2022, multiple observations were 
made towards it. The thermal emission and absorption are weak but they do not 
appear to change with time, even at 4.766\,GHz when the maser is flaring. 
This allows us to combine all the observations so that spectra with a better 
SNR are obtained. Most of the plots shown in Figure~\ref{fig:G213_67} are an 
average of 18 spectra from observations taken on the 16 dates listed in 
Table~2 of \citetalias{SF25} as well as two observations made on 2021 January 
05 and November 27. Three 4.660\,GHz and one 6.017\,GHz observations were 
discarded due to radio frequency interference, and in seven sessions the 
6.017\,GHz frequency was replaced by 4.829\,GHz in order to look for 
formaldehyde. Gaussians fitted to these 4.7 and 6.0\,GHz data are listed in 
Appendix~\ref{ApC} Table~\ref{tab:MonR2}. 

\subsection{G189.030+0.783} 

There have been no 4.7\,GHz exOH masers reported in this source, but 6.035\,GHz 
masers have been found previously. \citet{AQF16} discovered a RCP maser at $v = 
3.27$\,\kms\ with a flux density $S_{\nu} = 0.98$\,Jy and an LCP component at $v 
= 3.36$\,\kms\ with $S_{\nu} = 0.6$\,Jy. If these are a Zeeman pair the Zeeman 
velocity is $\Delta V_\textrm{Z} = -45$\,\ms. A survey carried out using the 
Torum 32m telescope by \citet{SWB20} found an RCP peak at $v = 3.32$\,\kms\ with 
$S_{\nu} = 2.43$\,Jy and an LCP peak at $v = 3.37$\,\kms\ with $S_{\nu} = 
1.74$\,Jy, which they did not classify as a Zeeman pair. Using the Irbene 32m 
telescope with RCP and LCP feeds, \citet{PAS21} found an RCP flux density $S_{\nu} 
= 2.33$\,Jy at $v = 3.398$\,\kms\ and an LCP line with $S_{\nu} = 1.36$\,Jy at $v 
= 3.413$\,\kms. This gives a Zeeman velocity $\Delta V_\textrm{Z} = -7.5$\,\ms.

\begin{figure*}[ht!]
\centering
\scalebox{0.6}{\plotone{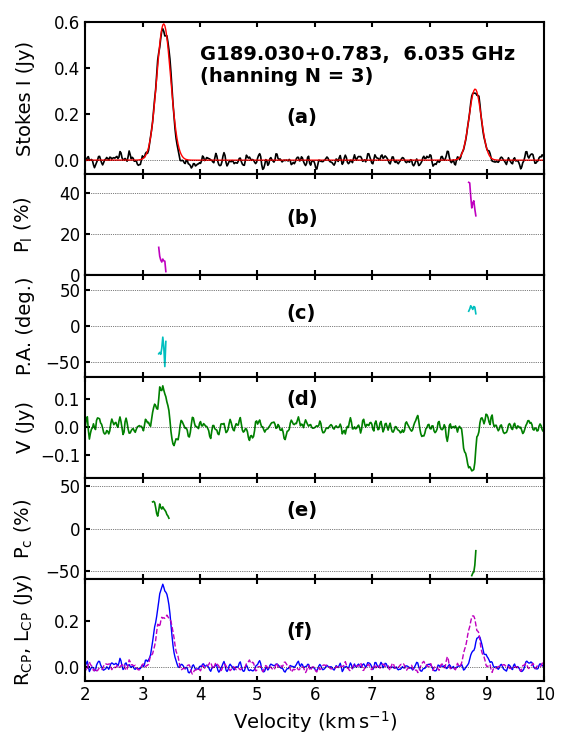}}
         \caption{Spectra of the 6.035\,GHz exOH in G189.030+0.783 showing
         (a) Stokes $I$ and fitted Gaussians, (b) $P_\textrm{l}$, (c) P.A.,
         (d) Stokes $V$, (e) $P_\textrm{c}$, and (f) the RCP and LCP signals.}
    \label{fig:G189_5}
\end{figure*}

We found two lines of 6.035\,GHz exOH masers towards this source, as can be 
seen in the spectra shown in Figure~\ref{fig:G189_5}, each of which was fitted 
with the single Gaussians listed in Table \ref{tab:GaussFit6}. These lines 
are narrow enough to each be single spots of emission, consistent with the 
residuals being noise. The peak flux density in the $v = 3.37$\,\kms\ component 
is weaker than in previous reports. The line at $v = 8.79$\,\kms, which has not 
been reported before, has a velocity close to that of the methanol masers in 
this source. There is linear and circular polarization in both spots, but, 
because of differences in the amplitudes of the circular components, there is 
no clear S-shape profile. The polarization is predominantly RCP for 
the 8.79\,\kms\ component, and LCP for the $v = 3.37$\,\kms\ component. In 
addition to the change in $P_\textrm{c}$ from positive to negative, $P_\textrm{l}$ 
changes from $\sim5$ to $\sim40$\% and the P.A. changes from $\sim -40$ to 
$+40$\degr\ for the $v = 8.79$ and 3.37\,\kms\ features. This indicates a 
shift in the orientation of the magnetic field between the two masering regions.
Zeeman splitting in these lines is discussed in section \ref{sec:mag}.

\subsection{G188.946+0.886 (IRAS 06058+2138)} 
There are no previous reports of 4.7 or 6.0\,GHz exOH masers in this source.

\begin{figure*}[ht!]
\centering
\scalebox{0.6}{\plotone{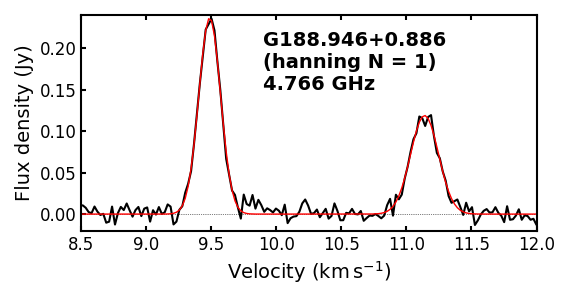}}
        \caption{Stokes $I$ spectrum with fitted Gaussians of the new detection 
        of 4.766\,GHz exOH in G188.946+0.886.}
    \label{fig:G188_0}
\end{figure*}

Our 4.766\,GHz Stokes $I$ spectrum of G188.030+0.783 is presented in Figure 
\ref{fig:G188_0}, showing the detection of two new peaks whose Gaussian fitted 
parameters are listed in Table \ref{tab:GaussFit47}. The lines are weak and 
have narrow widths, suggesting that these are two new isolated maser spots. No 
polarization was detected above the noise which means that if there is any $Q$, 
$U$ or $V$ signal it is less than 10\% of the $I$ signal. 

\subsection{G196.454--1.677} 

There are no reports in the literature of any 4.7 or 6.0\,GHz exOH masers in 
this source. 

\begin{figure*}[ht!]
\centering
         \scalebox{0.6}{\plotone{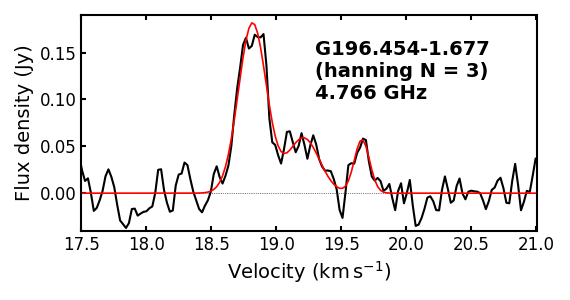}}
        \caption{Stokes $I$ spectrum with fitted Gaussians of the new detection 
        of 4.766\,GHz exOH in G196.454--1.677.}
        \label{fig:G196_0}
\end{figure*}

We report a new discovery of 4.766\,GHz masers, the $I$ spectrum of which is 
shown in Figure~\ref{fig:G196_0}. Gaussians have been fitted to three features 
in the spectrum and are listed in Table~\ref{tab:GaussFit47}. No polarization 
was detected in these weak maser lines. The 6.7\,GHz masers in this source are 
rated as being highly variable \citep{ASS23}, so it will be interesting to see 
if these 4.7\,GHz exOH display similar, highly variable characteristics.

\subsection{G240.316+0.071 (IRAS 07427--2400)} 

\citet{DE02} reported 4.766\,GHz exOH masers with peak flux densities of 0.17 
and 0.31\,Jy at velocities $v = 62.9$ and 65.0\,\kms, respectively. Subsequent 
searches have not found these masers, suggesting that they were short-lived. 

\citet{S94} discovered 6.035\,GHz exOH RCP and LCP masers with a velocity 
$v = 63.605$\,\kms\ and peak flux densities of 0.8 and 2.2\,Jy, respectively, 
showing no detectable Zeeman splitting. The higher sensitivity observations 
of \citet{CV95} found a second much weaker line with a flux density of 
$~0.2$\,Jy showing equal RCP and LCP components at $v = 62.7$\,\kms.  Further 
observations made by \citet{C03} and \citet{AQF16} found the same lines at the
same velocities. The stability of this maser was confirmed by \citet{QOS25} who 
identified three maser lines at $v = 62.82, 63.66$ and 63.99\,\kms. The LCP line
is more than three times stronger than the RCP line, as noted previously. The 
feature at $v = 62.82$\,\kms\ had a Zeeman splitting due to a field $B_{\parallel} 
= 0.3$\,mG. 

\begin{figure*}[ht!]
\centering
  \scalebox{0.6}{\plotone{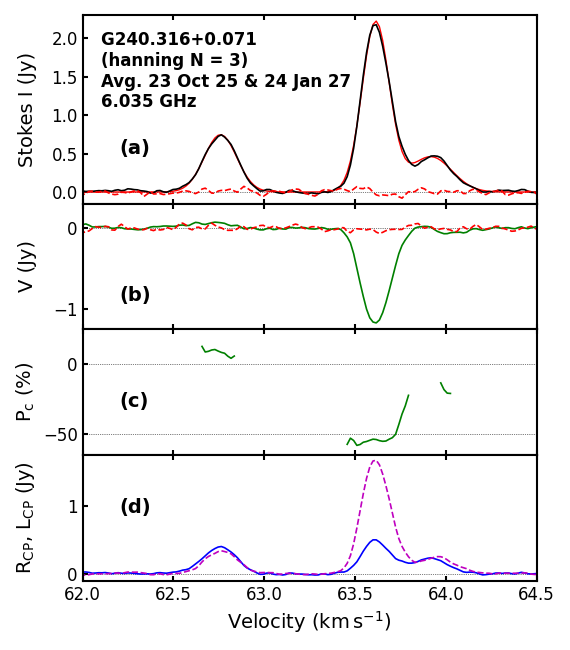}}
    \caption{Averaged plots of the two 6.035\,GHz observations towards 
    G240.316+0.071 showing (a) Stokes $I$, (b) Stokes $V$, (c) $P_\textrm{c}$, 
    and (d) RCP \& LCP. The dashed red lines in (a) and (b) are the difference 
    between the observations on 2023 October 25 and 2024 January 27, showing 
    that difference is at the level of the rms noise.}
    \label{fig:G240_5}
\end{figure*}

We got two sets of observations on G240.316+0.071, which showed no 4.7\,GHz 
sources and three lines of 6.035\,GHz exOH masers. The 6.035\,GHz spectra 
did not differ significantly between the two sets of observations. The 
Gaussian parameters of the fitted curves are listed in 
Table~\ref{tab:GaussFit6} and average of the two spectra are shown in 
Figure~\ref{fig:G240_5}.  

The strongest 6.035\,GHz peak at $v = 63.6$\,\kms\ has $\sim60$\% LCP, while 
the peaks at $v = 62.7$ and 63.9\,\kms\ show $\sim 10\%$ circular polarization 
in RCP and LCP respectively. No significant Stokes $Q$ or $U$ emission is 
present. These results are similar to those found previously, indicating that 
the masers in this source have been stable for many years.

\section{Zeeman splitting and Magnetic fields} \label{sec:mag}
\begin{figure*}[ht!]
\centering
         \scalebox{0.7}{\plotone{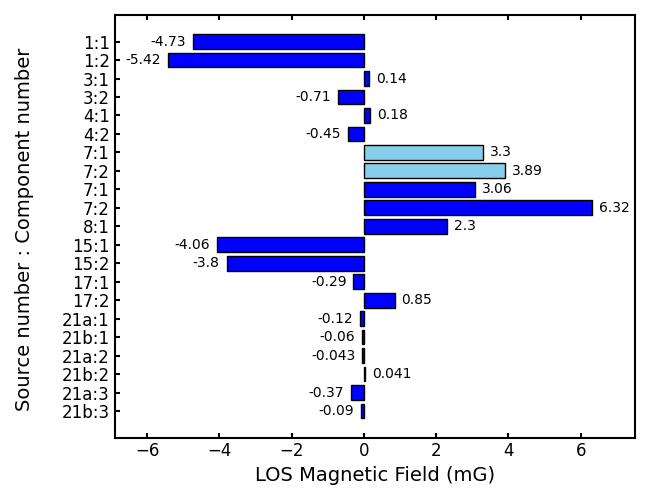}}
        \caption{Histogram of the $B_{\parallel}$ magnetic fields in 
        Table~\ref{tab:Mag_field}. Light blue is for 6.031\,GHz and dark
        blue for 6.035\,GHz.}
        \label{fig:hist}
\end{figure*}

In order to determine values of Zeeman splitting $\Delta V_\textrm{Z} = 
(v_\textrm{RCP} - v_\textrm{LCP})/2$, the velocities of the Gaussian 
fitted RCP and LCP profiles listed in Table \ref{tab:GaussFit6} have 
been used. Degaussed velocities are calculated from the same values 
using $(v_\textrm{RCP} + v_\textrm{LCP})/2$. The magnetic field 
($B_{\parallel}$) along the observer's LOS is assumed to be the cause 
of the Zeeman splitting and has been determined from the RCP and LCP 
velocity separation using the Land\'{e} $g$-factors given by \citet{D74}. 
For the 6.031 and 6.035\,GHz lines the $\Delta V_\textrm{Z}/B$ values we 
have used are 0.0790 and 0.0564\,\kms\,mG$^{-1}$. The results of these 
measurements and calculations are presented in Table~\ref{tab:Mag_field}, 
and a graphical representation of these $B$ fields is presented in 
Figure~\ref{fig:hist}.

\begin{deluxetable*}{ccccDDD}
\tablecaption{Gaussian parameters, Zeeman splitting and magnetic field values 
          from distinct profiles with apparent single masers.}
\label{tab:Mag_field}
\tablehead{
\colhead{No.} &\colhead{Source} &\colhead{Frequency} &\colhead{Component}      
       &\multicolumn2c{Degaussed}    &\multicolumn2c{Zeeman}    &\multicolumn2c{Magnetic} \\
\colhead{}   &\colhead{Name}    &\colhead{}      &\colhead{No. from}   
       &\multicolumn2c{Velocity}            &\multicolumn2c{splitting} &\multicolumn2c{field} \\
\colhead{}    &\colhead{}       &\colhead{(GHz)}  &\colhead{Table \ref{tab:GaussFit6} }             &\multicolumn2c{(\kms)}    &\multicolumn2c{(\ms)}     &\multicolumn2c{(mG)}
    }
\decimalcolnumbers
\startdata
1  &G108.758-0.986  &6.035  &1   &-45.672(2)     &-267.(2)   &-4.73(4) \\
   &                &       &2   &-45.231(2)     &-306.(2)   &-5.42(4) \\
\hline
3 &G111.532+0.759   &6.035  &1   &-59.5053(17)   &8.0(18)    &0.14(3) \\
  &                 &       &2   &-59.150(9)     &-40.(9)    &-0.71(16) \\
\hline
4 &G111.542+0.777   &6.035  &1   &-59.5153(8)    &10.2(8)    &0.180(15) \\
  &                 &       &2   &-59.148(3)     &-25.(3)    &-0.45(5) \\
\hline
7 &G133.947+1.064   &6.031  &1   &-43.206(13)    &259.(13)   &3.30(17) \\
  &                 &       &2   &-43.034(4)     &306.(4)    &3.89(5) \\
\\
  &                 &6.035  &1   &-43.161(3)     &172.(3)     &3.06(5) \\
  &                 &       &2   &-42.505(6)     &355.(6)     &6.32(10) \\
\hline
8 &G141.918+1.902   &6.035  &1   &-7.975(11)     &129.(11)    &2.3(2) \\
\hline
15 &G183.349–0.575  &6.035  &1   &-5.515(9)      &-0.229(9)   &-4.06(16) \\
   &                &       &2   &-5.095(12)     &-0.216(11)  &-3.8(2) \\  
\hline
17 &G189.030+0.783  &6.035  &1   &3.370(2)       &-16.(2)    &-0.29(4) \\
   &                &       &2   &8.806(3)       &48.(3)     &0.85(6) \\
\hline
21a &G240.316+0.071 &6.035  &1   &62.7637(15)    &-6.6(15)  &-0.12(3) \\ 
21b &               &       &1   &62.7554(15)    &-3.6(15)  &-0.06(3) \\
\\
21a &               &       &2   &63.6140(11)    &-2.4(11)  &-0.043(19) \\ 
21b &               &       &2   &63.6138(10)    &2.3(10)   &0.041(17) \\
\\
21a &               &       &3   &63.921(3)      &-21.(3)   &-0.37(6) \\ 
21b &               &       &3   &63.916(4)      &-5.(4)    &-0.09(6) \\
\hline
\enddata
\end{deluxetable*}

In G108.758–-0.986, the 6.035\,GHz RCP and LCP spectra are separated with each 
having two profiles. We have assumed that the lower velocity profiles are a Zeeman 
pair and the higher velocity components are a second pair. This is in agreement 
with the interferometric observations of \citet{KBR25}, who obtained similar 
degaussed velocities and magnetic field orientation to ours. Our values of 
$B_{\parallel}$ are about half their values because we have taken the Zeeman 
splitting to be half the difference between the velocity separation. Small 
differences in the value of the magnetic fields between their observations and 
ours could be due to variations in $B$, a feature that we have seen in some 
methanol masers in star-forming regions \citepalias{FS26}. \citet{OCS22} 
report a single value of $B_{\parallel} = -5.2$\,mG at a degaussed velocity 
$v = -45.81$\,\kms, that aligns with the two values we observe at slightly 
lower degaussed velocities.

Of our four pointings towards NGC7538, only two of them (observations 3 and 4 
in Table \ref{tab:source}) had sufficient SNR to detect polarization. These 
two observations are separated by about a month. Previous observations made 
years apart (see section \ref{sec:7538}) found peaks with different velocities 
each time this source was studied, indicating a variable source. The Zeeman 
splittings and magnetic fields determined from our spectra show differences on 
the limit of the uncertainties, suggesting that these variations are real. 
Interestingly, the directions of the magnetic field determined from the two sets 
of Zeeman pairs show a field pointing towards us for the redder pair, and away 
from us for the bluer pair. \citet{OCS22} also found this but for different 
degaussed velocities. The changes are indicative of a varying magnetic field 
and warrant further study on time scales of weeks to months.

W3(OH) is the only source in our sample that has measureable Zeeman components 
at both 6.031 and 6.035\,GHz. These complex spectra consist of numerous 
overlapping spots of emission identified in interferometric observations, so 
the features we have identified are somewhat tentative, and features could have 
varied since the earlier observations so direct comparisons are not possible. 
However, the values we obtain for $B_{\parallel}$ are in the range found 
previously and have the same orientation.

The G141.918+1.902 RCP and LCP spectra have very different amplitudes at 
6.035\,GHz, but are separated by a small velocity. Assuming the velocity 
separation is due to Zeeman splitting, allows a calculation of the magnetic 
field. We find a field of similar magnitude, but opposite direction, to the 
6.031\,GHz Zeeman splitting calculations of \citet{OCS22}. The strengths of
the masers have changed between their observations and ours, and it is 
possible that the magnetic field is also varying, as we suggested for 
NGC7538 and G240.316+0.071, and have shown for some 6.7\,GHz methanol masers 
in \citetalias{FS26}. Clearly, more follow-up observations are required.

RCP and LCP spectra in G183.349–0.575 each have two profiles, suggesting 
two sets of Zeeman pairs. We have assumed the lower velocity profiles are 
a one pair and the higher velocity components a second pair. The two 
calculated magnetic fields have similar magnitudes to other $B$ field 
values we report here. \citet{OCS22} identify a Zeeman pair but at 
different velocities to ours, and reported $B_{\parallel} = -6.4$\,mG. 
This is higher than our value but has the same orientation that we get. 

A weak magnetic field has been reported previously in G189.030+0.783 for the 
line at $v = 3.37$\,\kms, which we also find with the same orientation as  
the earlier observations. The new feature at $v = 8.81$\,\kms\ has a stronger 
$B$ field than the 3.37\,\kms\ line, and an orientation in the opposite direction. 
Methanol spectra \citepalias{FS26} displayed small degrees of both linear and 
circular polarization, but no Zeeman splitting was found, which is not 
surprising given how weak this $B$ field is.

We made two observations towards G240.316+0.071 separated by 3 months. Although 
the two sets of spectra looked the same, there are some subtle differences in 
the fitted line parameters that are smaller than the velocity resolution of the
spectra. The differences between the fitted parameters in Table \ref{tab:Mag_field} 
for the 21a and 21b epochs are larger than the uncertainties in most cases, 
indicating that the magnetic field in this region could be changing. Clearly,
more monitoring observations are needed to confirm this.

\section{Conclusions}
New 4.766\,GHz exOH masers are reported in G188.946+0.886 and G196.454--1.677. 
No new sources of 4.7\,GHz linear or circular polarization have been found, 
nor have any new 4.751 or 4.660\,GHz masers been discovered. Lines differing 
from those reported previously were found in G133.715+1.215. 

New 6.035\,GHz exOH masers were found in G240.316+0.071. No new 6.031\,GHz 
exOH masers are reported, nor are there any 6.017 or 6.049\,GHz new exOH 
masers. We observed a 6.049\,GHz feature in G133.947+1.064, and as we have 
higher resolution spectra than previously recorded spectra, we can resolve 
this emission into a broad thermal component and a maser feature which is 
quite broad (0.40\,\kms). 

We found a number of thermal emission and absorption lines that have not been 
reported before. Thermal emission at 4.766\,GHz and absorption at 6.017, 6.031 
and 6.035\,GHz was found in G133.715+1.215. Thermal emission at 4.751 and 
6.049\,GHz was identified in G133.947+1.064, and an absorption line at 4.660\,GHz. 
In Mon R2 we found thermal emission at 4.751 and 4.766\,GHz that have a fitted 
Gaussian peak at the same velocity, and absorption at 4.660\,GHz at a slightly 
(statistically significant) lower velocity, neither of which coincide with the
velocity of the maser profile. Absorption was detected at 6.017, 6.031 and 6.035\,GHz 
with dips occurring at velocities different to those of the 4.766\,GHz flaring 
exOH maser and the 6.7,GHz methanol masers. The 6.031 and 6.035\,GHz profiles 
are asymmetrical, suggesting more than one velocity component contributing to
the overall shape.

The 6.035\,GHz masers in NGC7538, G133.947+1.064 and G189.030+0.78, and the 
6.031\,GHz masers in G133.947+1.064 are the four sources in our sample that
have linear polarized components. In addition, linear polarization changes 
between epochs in the 4.766\,GHz Mon R2 maser and the 6.035\,GHz NGC7538 
masers are found. In both instances, there has been a shift in the orientation 
of the magnetic field over time. 

Zeeman splittings have been calculated for all 6.0\,GHz exOH masers that 
display circular polarization, however, our assumptions in determining 
Zeeman pairs requires interferometric confirmation. There are indications 
from these observations of magnetic fields varying on time scales of weeks 
to months. 

\begin{acknowledgments}
This material is based upon work supported by the National Radio Astronomy 
Observatory and Green Bank Observatory which are major facilities funded 
by the U.S. National Science Foundation operated by Associated Universities, 
Inc. The observations were part of GBT project AGBT22B-354. This research has 
made use of the Science Explorer, funded by NASA under Cooperative Agreement 
80NSSC21M00561. This research has made use of the SIMBAD database, operated 
at CDS, Strasbourg, France
\added

\end{acknowledgments}

\begin{contribution}
DPS did the literature reviews and wrote most of the text. PF did the observations 
and data reductions, produced the graphs, and contributed to editing the document.
\end{contribution}

\vspace{5mm}
\facilities{GBT:100m}

\software{GBTIDL \citep{MGB13}, Matplotlib \citep{Hunter:2007}, MaserDB  \citep{LBS19}}

\bibliographystyle{aasjournalv7}
\bibliography{AAS_exOH}

@ARTICLE{ZPP68,
       author = {{Zuckerman}, B. and {Palmer}, Patrick and {Penfield}, H. and {Lilley}, A.~E.},
        title = "{Detection of Microwave Radiation from the \^\{2\}{\ensuremath{\Pi}}\_\{{\textonehalf}\}, J = {\textonehalf} State of OH}",
      journal = {\apjl},
         year = 1968,
        month = jul,
       volume = {153},
        pages = {L69},
          doi = {10.1086/180223},
       adsurl = {https://ui.adsabs.harvard.edu/abs/1968ApJ...153L..69Z}
}

@ARTICLE{YZP69,
       author = {{Yen}, J.~L. and {Zuckerman}, B. and {Palmer}, Patrick and {Penfield}, H.},
        title = "{Detection of the $^{2}${\ensuremath{\Pi}}$_{3/2}$, J = 5/2; State of OH at 5-centimeter Wavelength}",
      journal = {\apjl},
         year = 1969,
        month = apr,
       volume = {156},
        pages = {L27},
          doi = {10.1086/180342},
       adsurl = {https://ui.adsabs.harvard.edu/abs/1969ApJ...156L..27Y}
}

@ARTICLE{ZP70,
       author = {{Zuckerman}, B. and {Palmer}, Patrick},
        title = "{Observations of the \^\{2\}{\ensuremath{\Pi}}\_\{{\textonehalf}\}, J = {\textonehalf} State of Interstellar OH}",
      journal = {\apjl},
         year = 1970,
        month = mar,
       volume = {159},
        pages = {L197},
          doi = {10.1086/180510},
       adsurl = {https://ui.adsabs.harvard.edu/abs/1970ApJ...159L.197Z}
}

@ARTICLE{TWB70,
       author = {{Thacker}, D.~L. and {Wilson}, W.~J. and {Barrett}, A.~H.},
        title = "{Observations of the $^{2}\Pi\_{1/2}$, J = 1/2 State of OH}",
      journal = {\apjl},
         year = 1970,
        month = sep,
       volume = {161},
        pages = {L191},
          doi = {10.1086/180601},
       adsurl = {https://ui.adsabs.harvard.edu/abs/1970ApJ...161L.191T}
}

@ARTICLE{ZYGP72,
       author = {{Zuckerman}, B. and {Yen}, J.~L. and {Gottlieb}, C.~A. and {Palmer}, Patrick},
        title = "{Observations of the $^{2}\Pi_{3/2}$, J = 5/2 State of Interstellar OH}",
      journal = {\apj},
         year = 1972,
        month = oct,
       volume = {177},
        pages = {59},
          doi = {10.1086/151686},
       adsurl = {https://ui.adsabs.harvard.edu/abs/1972ApJ...177...59Z}
}

@INPROCEEDINGS{D74,
       author = {{Davies}, R.~D.},
        title = "{Magnetic Fields in OH Maser Clouds}",
    booktitle = {Galactic Radio Astronomy},
         year = 1974,
       editor = {{Kerr}, Frank J. and {Simonson}, Simon Christian},
       series = {IAU Symposium},
       volume = {60},
        month = jan,
        pages = {275},
       adsurl = {https://ui.adsabs.harvard.edu/abs/1974IAUS...60..275D}
}

@ARTICLE{B74,
       author = {{Baudry}, A.},
        title = "{Observations of the Excited Lines of OIL at a Wavelength of 6.3 CM}",
      journal = {\aap},
         year = 1974,
        month = jul,
       volume = {33},
        pages = {381},
       adsurl = {https://ui.adsabs.harvard.edu/abs/1974A&A....33..381B}
}

@ARTICLE{RZP75,
       author = {{Rickard}, L.~J. and {Zuckerman}, B. and {Palmer}, P.},
        title = "{Observations of quasi-thermal and maser phenomena in rotationally excited OH. I.}",
      journal = {\apj},
         year = 1975,
        month = aug,
       volume = {200},
        pages = {6-21},
          doi = {10.1086/153755},
       adsurl = {https://ui.adsabs.harvard.edu/abs/1975ApJ...200....6R}
}

@ARTICLE{MRL78,
       author = {{Moran}, J.~M. and {Reid}, M.~J. and {Lada}, C.~J. and {Yen}, J.~L. and {Johnston}, K.~J. and {Spencer}, J.~H.},
        title = "{Evidence for the Zeeman effect in the OH maser emission from W3(OH).}",
      journal = {\apjl},
         year = 1978,
        month = sep,
       volume = {224},
        pages = {L67-L71},
          doi = {10.1086/182761},
       adsurl = {https://ui.adsabs.harvard.edu/abs/1978ApJ...224L..67M}
}

@ARTICLE{GMP83,
       author = {{Gardner}, F.~F. and {Martin-Pintado}, J.},
        title = "{Emission and absorption at 6 CM from excited OH associated with compact HII regions}",
      journal = {\aap},
         year = 1983,
        month = may,
       volume = {121},
        pages = {265-270},
       adsurl = {https://ui.adsabs.harvard.edu/abs/1983A&A...121..265G}
}

@ARTICLE{GWP83,
       author = {{Gardner}, F.~F. and {Whiteoak}, J.~B. and {Palmer}, P.},
        title = "{OH maser emission at 4765 MHz in W 3.}",
      journal = {\mnras},
         year = 1983,
        month = oct,
       volume = {205},
        pages = {297-302},
          doi = {10.1093/mnras/205.2.297},
       adsurl = {https://ui.adsabs.harvard.edu/abs/1983MNRAS.205..297G}
}

@ARTICLE{GBW84,
       author = {{Guilloteau}, S. and {Baudry}, A. and {Walmsley}, C.~M. and {Wilson}, T.~L. and {Winnberg}, A.},
        title = "{Rotationally excited OH : emission and absorption toward HII/OH regions.}",
      journal = {\aap},
         year = 1984,
        month = feb,
       volume = {131},
        pages = {45-57},
       adsurl = {https://ui.adsabs.harvard.edu/abs/1984A&A...131...45G}
}

@ARTICLE{PGW84,
       author = {{Palmer}, P. and {Gardner}, F.~F. and {Whiteoak}, J.~B.},
        title = "{OH maser emission at 4765 MHz in NGC 7538 (IRS1).}",
      journal = {\mnras},
         year = 1984,
        month = nov,
       volume = {211},
        pages = {41P-44},
          doi = {10.1093/mnras/211.1.41P},
       adsurl = {https://ui.adsabs.harvard.edu/abs/1984MNRAS.211P..41P}
}

@ARTICLE{CMW91,
       author = {{Cohen}, R.~J. and {Masheder}, M.~R.~W. and {Walker}, R.~N.~F.},
        title = "{Excited OH 4.7-GHz masers associated with IRAS far-intrared sources.}",
      journal = {\mnras},
         year = 1991,
        month = jun,
       volume = {250},
        pages = {611},
          doi = {10.1093/mnras/250.3.611},
       adsurl = {https://ui.adsabs.harvard.edu/abs/1991MNRAS.250..611C}
}

@BOOK{E92,
       author = {{Elitzur}, Moshe},
        title = "{Astronomical masers}",
         year = 1992,
       volume = {170},
    publisher = {Springer Dordrecht},
          doi = {10.1007/978-94-011-2394-5},
       adsurl = {https://ui.adsabs.harvard.edu/abs/1992ASSL..170.....E}
}

@ARTICLE{S94,
       author = {{Smits}, D.~P.},
        title = "{New detections of excited OH masers at 5-cm wavelength.}",
      journal = {\mnras},
         year = 1994,
        month = jul,
       volume = {269},
        pages = {L11-L16},
          doi = {10.1093/mnras/269.1.L11},
       adsurl = {https://ui.adsabs.harvard.edu/abs/1994MNRAS.269L..11S}
}

@ARTICLE{CV95,
       author = {{Caswell}, J.~L. and {Vaile}, R.~A.},
        title = "{Excited-state OH masers at 6.035 GHz}",
      journal = {\mnras},
         year = 1995,
        month = mar,
       volume = {273},
       number = {2},
        pages = {328-346},
          doi = {10.1093/mnras/273.2.328},
       adsurl = {https://ui.adsabs.harvard.edu/abs/1995MNRAS.273..328C}
}

@ARTICLE{CMC95,
       author = {{Cohen}, R.~J. and {Masheder}, M.~R.~W. and {Caswell}, J.~L.},
        title = "{Excited OH 4.7-GHz masers associated with IRAS far-infrared sources - II}",
      journal = {\mnras},
         year = 1995,
        month = jun,
       volume = {274},
       number = {3},
        pages = {808-820},
          doi = {10.1093/mnras/274.3.808},
       adsurl = {https://ui.adsabs.harvard.edu/abs/1995MNRAS.274..808C}
}

@ARTICLE{M97,
       author = {{MacLeod}, G.~C.},
        title = "{1720-MHz OH masers: distribution and associations with excited-OH masers}",
      journal = {\mnras},
         year = 1997,
        month = mar,
       volume = {285},
       number = {3},
        pages = {635-639},
          doi = {10.1093/mnras/285.3.635},
       adsurl = {https://ui.adsabs.harvard.edu/abs/1997MNRAS.285..635M}
}

@ARTICLE{S97,
       author = {{Smits}, Derck P.},
        title = "{Monitoring of 6-cm excited OH masers}",
      journal = {\mnras},
         year = 1997,
        month = may,
       volume = {287},
       number = {2},
        pages = {253-261},
          doi = {10.1093/mnras/287.2.253},
       adsurl = {https://ui.adsabs.harvard.edu/abs/1997MNRAS.287..253S}
}

@ARTICLE{BDW97,
       author = {{Baudry}, A. and {Desmurs}, J.~F. and {Wilson}, T.~L. and {Cohen}, R.~J.},
        title = "{A survey of star-forming regions in the 5 CM lines of OH.}",
      journal = {\aap},
         year = 1997,
        month = sep,
       volume = {325},
        pages = {255-268},
       adsurl = {https://ui.adsabs.harvard.edu/abs/1997A&A...325..255B}
}

@ARTICLE{SCH98,
       author = {{Smits}, Derck P. and {Cohen}, R.~J. and {Hutawarakorn}, B.},
        title = "{OH 4765-MHz maser flares in MON R2}",
      journal = {\mnras},
         year = 1998,
        month = may,
       volume = {296},
       number = {1},
        pages = {L11-l14},
          doi = {10.1046/j.1365-8711.1998.01515.x},
       adsurl = {https://ui.adsabs.harvard.edu/abs/1998MNRAS.296L..11S}
}

@ARTICLE{SKH00,
       author = {{Szymczak}, M. and {Kus}, A.~J. and {Hrynek}, G.},
        title = "{Observations of OH 4765-MHz maser emission from star-forming regions}",
      journal = {\mnras},
         year = 2000,
        month = feb,
       volume = {312},
       number = {1},
        pages = {211-216},
          doi = {10.1046/j.1365-8711.2000.03214.x},
       adsurl = {https://ui.adsabs.harvard.edu/abs/2000MNRAS.312..211S}
}

@ARTICLE{DE02,
       author = {{Dodson}, R.~G. and {Ellingsen}, S.~P.},
        title = "{A search for 4750- and 4765-MHz OH masers in southern star-forming regions}",
      journal = {\mnras},
         year = 2002,
        month = jun,
       volume = {333},
       number = {2},
        pages = {307-317},
          doi = {10.1046/j.1365-8711.2002.05428.x},
archivePrefix = {arXiv},
       eprint = {astro-ph/0201296},
 primaryClass = {astro-ph},
       adsurl = {https://ui.adsabs.harvard.edu/abs/2002MNRAS.333..307D}
}

@ARTICLE{S03,
       author = {{Smits}, Derck P.},
        title = "{Monitoring of 6-cm excited OH masers - II}",
      journal = {\mnras},
         year = 2003,
        month = feb,
       volume = {339},
       number = {1},
        pages = {1-11},
          doi = {10.1046/j.1365-8711.2003.06096.x},
       adsurl = {https://ui.adsabs.harvard.edu/abs/2003MNRAS.339....1S}
}

@ARTICLE{C03,
       author = {{Caswell}, J.~L.},
        title = "{Spectra of OH masers at 6035 and 6030 MHz}",
      journal = {\mnras},
         year = 2003,
        month = may,
       volume = {341},
       number = {2},
        pages = {551-568},
          doi = {10.1046/j.1365-8711.2003.06418.x},
       adsurl = {https://ui.adsabs.harvard.edu/abs/2003MNRAS.341..551C}
}

@article{PGW04,
	author = {Palmer, Patrick and Goss, W. M. and Whiteoak, J. B.},
	title = "{VLA observations of 6-cm excited OH}",
	journal = {\mnras},
	volume = {347},
	number = {4},
	pages = {1164-1172},
	year = {2004},
	month = {Feb},
	issn = {0035-8711},
	doi = {10.1111/j.1365-2966.2004.07280.x},
	url = {https://doi.org/10.1111/j.1365-2966.2004.07280.x},
	eprint = {https://academic.oup.com/mnras/article-pdf/347/4/1164/3192812/347-4-1164.pdf},
}

@ARTICLE{OTN04,
       author = {{Ojha}, D.~K. and {Tamura}, M. and {Nakajima}, Y. and {Fukagawa}, M. and {Sugitani}, K. and {Nagashima}, C. and {Nagayama}, T. and {Nagata}, T. and {Sato}, S. and {Vig}, S. and {Ghosh}, S.~K. and {Pickles}, A.~J. and {Momose}, M. and {Ogura}, K.},
        title = "{A Near-Infrared Study of the NGC 7538 Star-forming Region}",
      journal = {\apj},
         year = 2004,
        month = dec,
       volume = {616},
       number = {2},
        pages = {1042-1057},
          doi = {10.1086/425068},
archivePrefix = {arXiv},
       eprint = {astro-ph/0408219},
 primaryClass = {astro-ph},
       adsurl = {https://ui.adsabs.harvard.edu/abs/2004ApJ...616.1042O}
}

@ARTICLE{HSC05,
       author = {{Harvey-Smith}, L. and {Cohen}, R.~J.},
        title = "{A MERLIN survey of 4.7-GHz excited OH masers in star-forming regions}",
      journal = {\mnras},
         year = 2005,
        month = jan,
       volume = {356},
       number = {2},
        pages = {637-646},
          doi = {10.1111/j.1365-2966.2004.08485.x},
       adsurl = {https://ui.adsabs.harvard.edu/abs/2005MNRAS.356..637H}
}

@ARTICLE{ECG05,
       author = {{Etoka}, S. and {Cohen}, R.~J. and {Gray}, M.~D.},
        title = "{The association of OH and methanol masers in W3(OH)}",
      journal = {\mnras},
         year = 2005,
        month = jul,
       volume = {360},
       number = {3},
        pages = {1162-1170},
          doi = {10.1111/j.1365-2966.2005.09130.x},
       adsurl = {https://ui.adsabs.harvard.edu/abs/2005MNRAS.360.1162E}
}

@ARTICLE{PMM06,
       author = {{Pestalozzi}, M.~R. and {Minier}, V. and {Motte}, F. and {Conway}, J.~E.},
        title = "{Discovery of two new methanol masers in NGC 7538. Locating of massive protostars}",
      journal = {\aap},
         year = 2006,
        month = mar,
       volume = {448},
       number = {3},
        pages = {L57-L60},
          doi = {10.1051/0004-6361:200600006},
       adsurl = {https://ui.adsabs.harvard.edu/abs/2006A&A...448L..57P}
}

@ARTICLE{FRM06,
       author = {{Fish}, V.~L. and {Reid}, M.~J. and {Menten}, K.~M. and {Pillai}, T.},
        title = "{Enhanced density and magnetic fields in interstellar OH masers}",
      journal = {\aap},
         year = 2006,
        month = nov,
       volume = {458},
       number = {2},
        pages = {485-495},
          doi = {10.1051/0004-6361:20065379},
archivePrefix = {arXiv},
       eprint = {astro-ph/0608121},
 primaryClass = {astro-ph},
       adsurl = {https://ui.adsabs.harvard.edu/abs/2006A&A...458..485F}
}

@ARTICLE{FZS06,
       author = {{Fish}, Vincent L. and {Zschaechner}, Laura K. and
         {Sjouwerman}, Lor{\'a}nt O. and {Pihlstr{\"o}m}, Ylva M. and
         {Claussen}, Mark J.},
        title = "{Observations of the 6 cm Lines of OH in Evolved (OH/IR) Stars}",
      journal = {\apjl},
         year = 2006,
        month = dec,
       volume = {653},
       number = {1},
        pages = {L45-L48},
          doi = {10.1086/510382},
archivePrefix = {arXiv},
       eprint = {astro-ph/0610709},
 primaryClass = {astro-ph},
       adsurl = {https://ui.adsabs.harvard.edu/abs/2006ApJ...653L..45F}
}

@Article{Hunter:2007,
  Author    = {Hunter, J. D.},
  Title     = {Matplotlib: A 2D graphics environment},
  Journal   = {Computing in Science \& Engineering},
  Volume    = {9},
  Number    = {3},
  Pages     = {90--95},
  publisher = {IEEE COMPUTER SOC},
  doi       = {10.1109/MCSE.2007.55},
  year      = 2007
}

@ARTICLE{FSP07,
       author = {{Fish}, Vincent L. and {Sjouwerman}, Lor{\'a}nt O. and {Pihlstr{\"o}m}, Ylva M.},
        title = "{Effelsberg Observations of Excited-State (6.0 GHz) OH in Supernova Remnants and W3(OH)}",
      journal = {\apjl},
         year = 2007,
        month = dec,
       volume = {670},
       number = {2},
        pages = {L117-L120},
          doi = {10.1086/524196},
archivePrefix = {arXiv},
       eprint = {0710.1770},
 primaryClass = {astro-ph},
       adsurl = {https://ui.adsabs.harvard.edu/abs/2007ApJ...670L.117F}
}

@ARTICLE{GMM09,
       author = {{Gutermuth}, R.~A. and {Megeath}, S.~T. and {Myers}, P.~C. and {Allen}, L.~E. and {Pipher}, J.~L. and {Fazio}, G.~G.},
        title = "{A Spitzer Survey of Young Stellar Clusters Within One Kiloparsec of the Sun: Cluster Core Extraction and Basic Structural Analysis}",
      journal = {\apjs},
         year = 2009,
        month = sep,
       volume = {184},
       number = {1},
        pages = {18-83},
          doi = {10.1088/0067-0049/184/1/18},
archivePrefix = {arXiv},
       eprint = {0906.0201},
 primaryClass = {astro-ph.SR},
       adsurl = {https://ui.adsabs.harvard.edu/abs/2009ApJS..184...18G}
}

@article{MGB13,
       author = {{Marganian}, P. and {Garwood}, R.~W. and {Braatz}, J.~A. and {Radziwill}, N.~M. and 
{Maddalena}, R.~J.},
        title = "{GBTIDL: Reduction and Analysis of GBT Spectral Line Data}",
 howpublished = {Astrophysics Source Code Library, record ascl:1303.019},
         year = 2013,
        month = mar,
      journal = {Astrophysics Source Code Library},
          eid = {ascl:1303.019},
       adsurl = {https://ui.adsabs.harvard.edu/abs/2013ascl.soft03019M}
}

@ARTICLE{AQF16,
       author = {{Avison}, A. and {Quinn}, L.~J. and {Fuller}, G.~A. and {Caswell}, J.~L. and {Green}, J.~A. and {Breen}, S.~L. and {Ellingsen}, S.~P. and {Gray}, M.~D. and {Pestalozzi}, M. and {Thompson}, M.~A. and {Voronkov}, M.~A.},
        title = "{Excited-state hydroxyl maser catalogue from the methanol multibeam survey - I. Positions and variability}",
      journal = {\mnras},
         year = 2016,
        month = may,
       volume = {461},
       number = {1},
        pages = {136-155},
          doi = {10.1093/mnras/stw1101},
archivePrefix = {arXiv},
       eprint = {1605.02506},
 primaryClass = {astro-ph.GA},
       adsurl = {https://ui.adsabs.harvard.edu/abs/2016MNRAS.461..136A}
}

@ARTICLE{MSG18,
       author = {{MacLeod}, G.~C. and {Smits}, D.~P. and {Goedhart}, S. and {Hunter}, T.~R. and {Brogan}, C.~L. and {Chibueze}, J.~O. and {van den Heever}, S.~P. and {Thesner}, C.~J. and {Banda}, P.~J. and {Paulsen}, J.~D.},
        title = "{A masing event in NGC 6334I: contemporaneous flaring of hydroxyl, methanol, and water masers}",
      journal = {\mnras},
         year = 2018,
        month = jul,
       volume = {478},
       number = {1},
        pages = {1077-1092},
          doi = {10.1093/mnras/sty996},
archivePrefix = {arXiv},
       eprint = {1804.05308},
 primaryClass = {astro-ph.SR},
       adsurl = {https://ui.adsabs.harvard.edu/abs/2018MNRAS.478.1077M}
}

@ARTICLE{SXC18,
       author = {{Sun}, Yan and {Xu}, Ye and {Chen}, Xi and {Fang}, Min and {Henkel}, Christian and {Yang}, Ji and {Menten}, Karl M. and {Chen}, Xue-Peng and {Jiang}, Zhi-Bo},
        title = "{Discovery of H$_{2}$O, CH$_{3}$OH, and OH Masers in the Extreme Outer Galaxy}",
      journal = {\apj},
         year = 2018,
        month = dec,
       volume = {869},
       number = {2},
          eid = {148},
        pages = {148},
          doi = {10.3847/1538-4357/aaee86},
       adsurl = {https://ui.adsabs.harvard.edu/abs/2018ApJ...869..148S}
}

@ARTICLE{LBS19,
       author = {{Ladeyschikov}, Dmitry A. and {Bayandina}, Olga S. and {Sobolev}, Andrey M.},
        title = "{Online Database of Class I Methanol Masers}",
      journal = {\aj},
         year = 2019,
        month = dec,
       volume = {158},
       number = {6},
          eid = {233},
        pages = {233},
          doi = {10.3847/1538-3881/ab4b4c},
archivePrefix = {arXiv},
       eprint = {1911.04742},
 primaryClass = {astro-ph.IM},
       adsurl = {https://ui.adsabs.harvard.edu/abs/2019AJ....158..233L}
}

@ARTICLE{SWB20,
       author = {{Szymczak}, M. and {Wolak}, P. and {Bartkiewicz}, A. and {Aramowicz}, M. and {Durjasz}, M.},
        title = "{A search for the OH 6035 MHz line in high-mass star-forming regions}",
      journal = {\aap},
         year = 2020,
        month = oct,
       volume = {642},
          eid = {A145},
        pages = {A145},
          doi = {10.1051/0004-6361/202039009},
archivePrefix = {arXiv},
       eprint = {2009.06291},
 primaryClass = {astro-ph.GA},
       adsurl = {https://ui.adsabs.harvard.edu/abs/2020A&A...642A.145S}
}

@ARTICLE{PAS21,
       author = {{Patoka}, O. and {Antyufeyev}, O. and {Shmeld}, I. and {Bezrukovs}, V. and {Bleiders}, M. and {Orbidans}, A. and {Aberfelds}, A. and {Shulga}, V.},
        title = "{New ex-OH maser detections in the northern celestial hemisphere}",
      journal = {\aap},
         year = 2021,
        month = aug,
       volume = {652},
          eid = {A17},
        pages = {A17},
          doi = {10.1051/0004-6361/202037623},
archivePrefix = {arXiv},
       eprint = {2106.11585},
 primaryClass = {astro-ph.GA},
       adsurl = {https://ui.adsabs.harvard.edu/abs/2021A&A...652A..17P}
}

@INCOLLECTION{RH21,
       author = {{Robishaw}, Timothy and {Heiles}, Carl},
        title = "{The Measurement of Polarization in Radio Astronomy}",
    booktitle = {The WSPC Handbook of Astronomical Instrumentation, Volume 1: Radio Astronomical Instrumentation},
         year = 2021,
       editor = {{Wolszczan}, Alex},
        pages = {127-158},
    publisher = {World Scientific Publishing Co},
          doi = {10.1142/9789811203770_0006},
       adsurl = {https://ui.adsabs.harvard.edu/abs/2021hai1.book..127R}
}

@ARTICLE{SVL22,
       author = {{Surcis}, G. and {Vlemmings}, W.~H.~T. and {van Langevelde}, H.~J. and {Hutawarakorn Kramer}, B. and {Bartkiewicz}, A.},
        title = "{EVN observations of 6.7 GHz methanol maser polarization in massive star-forming regions. V. Completion of the flux-limited sample}",
      journal = {\aap},
         year = 2022,
        month = feb,
       volume = {658},
          eid = {A78},
        pages = {A78},
          doi = {10.1051/0004-6361/202142125},
archivePrefix = {arXiv},
       eprint = {2111.08023},
 primaryClass = {astro-ph.GA},
       adsurl = {https://ui.adsabs.harvard.edu/abs/2022A&A...658A..78S}
}

@ARTICLE{QSB22,
       author = {{Qiao}, Hai-Hua and {Shen}, Zhi-Qiang and {Breen}, Shari L. and {Yang}, Kai and {Chen}, Xi and {Li}, Juan},
        title = "{6 cm OH Masers in Northern Star Formation Regions}",
      journal = {\apj},
         year = 2022,
        month = apr,
       volume = {928},
       number = {2},
          eid = {129},
        pages = {129},
          doi = {10.3847/1538-4357/ac5820},
       adsurl = {https://ui.adsabs.harvard.edu/abs/2022ApJ...928..129Q}
}

@ARTICLE{OCS22,
       author = {{Ouyang}, Xu-Jia and {Chen}, Xi and {Shen}, Zhi-Qiang and {Li}, Bin and {Wu}, Ya-Jun and {Chen}, Hong-Ying and {Li}, Xiao-Qiong and {Yang}, Kai and {Song}, Shi-Min and {Qiao}, Hai-Hua},
        title = "{An Excited-state OH Maser Survey toward WISE Point Sources}",
      journal = {\apjs},
         year = 2022,
        month = jun,
       volume = {260},
       number = {2},
          eid = {51},
        pages = {51},
          doi = {10.3847/1538-4365/ac634c},
       adsurl = {https://ui.adsabs.harvard.edu/abs/2022ApJS..260...51O}
}

@ARTICLE{FSG23,
       author = {{Fallon}, Paul and {Smits}, Derck P. and {Ghosh}, Tapasi and {Salter}, Christopher J. and {Salas}, Pedro},
        title = "{Point Source C-band Mueller Matrices for the Green Bank Telescope}",
      journal = {\aj},
         year = 2023,
        month = jul,
       volume = {166},
       number = {1},
          eid = {26},
        pages = {26},
          doi = {10.3847/1538-3881/acd762},
archivePrefix = {arXiv},
       eprint = {2305.18055},
 primaryClass = {astro-ph.IM},
       adsurl = {https://ui.adsabs.harvard.edu/abs/2023AJ....166...26F}
}

@ARTICLE{ASS23,
       author = {{Aberfelds}, A. and {{\v{S}}teinbergs}, J. and {Shmeld}, I. and {Burns}, R.~A.},
        title = "{Five years of 6.7-GHz methanol maser monitoring with Irbene radio telescopes}",
      journal = {\mnras},
         year = 2023,
        month = dec,
       volume = {526},
       number = {4},
        pages = {5699-5714},
          doi = {10.1093/mnras/stad3158},
archivePrefix = {arXiv},
       eprint = {2310.08273},
 primaryClass = {astro-ph.GA},
       adsurl = {https://ui.adsabs.harvard.edu/abs/2023MNRAS.526.5699A}
}

@ARTICLE{QOS25,
       author = {{Qiao}, Hai-Hua and {Ouyang}, Xu-Jia and {Shen}, Zhi-Qiang and {Breen}, Shari L. and {Yang}, Kai and {Chen}, Xi and {Li}, Juan},
        title = "{5 cm OH Masers in Northern Star Formation Regions}",
      journal = {\aj},
         year = 2025,
        month = jan,
       volume = {169},
       number = {1},
          eid = {12},
        pages = {12},
          doi = {10.3847/1538-3881/ad9002},
       adsurl = {https://ui.adsabs.harvard.edu/abs/2025AJ....169...12Q}
}

@ARTICLE{KBR25,
       author = {{Kobak}, A. and {Bartkiewicz}, A. and {Rygl}, K.~L.~J. and {Richards}, A.~M.~S. and {Szymczak}, M. and {Wolak}, P.},
        title = "{Physical conditions around high-mass young star-forming objects via simultaneous observations of excited OH and methanol masers}",
      journal = {\aap},
         year = 2025,
        month = mar,
       volume = {695},
          eid = {A149},
        pages = {A149},
          doi = {10.1051/0004-6361/202452657},
archivePrefix = {arXiv},
       eprint = {2503.11379},
 primaryClass = {astro-ph.SR},
       adsurl = {https://ui.adsabs.harvard.edu/abs/2025A&A...695A.149K}
}

@ARTICLE{SF25,
       author = {{Smits}, Derck P. and {Fallon}, Paul},
        title = "{First Detection of Circular Polarization in 4.7 GHz Excited OH Masers}",
      journal = {\apj},
         year = 2025,
        month = sep,
       volume = {990},
       number = {2},
          eid = {193},
        pages = {193},
          doi = {10.3847/1538-4357/adf639},
archivePrefix = {arXiv},
       eprint = {2508.05450},
 primaryClass = {astro-ph.GA},
       adsurl = {https://ui.adsabs.harvard.edu/abs/2025ApJ...990..193S}
}

@ARTICLE{FS26,
       author = {{Fallon}, Paul and {Smits}, Derck P.},
        title = "{Polarization Characteristics of a sample of 6.7 GHz Masers}",
      journal = {\apj},
         year = 2026,
        month = mar,
       volume = {999},
       number = {2},
          eid = {193},
        pages = {193},
          doi = {10.3847/1538-4357/adf639},
archivePrefix = {arXiv},
       eprint = {2508.05450},
 primaryClass = {astro-ph.GA},
       adsurl = {https://ui.adsabs.harvard.edu/abs/2025ApJ...990..193S}
}

\appendix

\section{4.7 GHz exOH maser detections}
\startlongtable
\begin{deluxetable*}{ll}
	\tablecaption{Sources in star-forming in which 4.7\,GHz exOH masers have been found. 
    Columns contain (1) the source names, and (2) the reference to the first detection.}
	\label{tab:discover4.7}
\tablehead{ 
\colhead{Source}   &\colhead{Reference}  } 
\decimalcolnumbers
\startdata
\textbf{4766} \\
W3(OH), W49             &\citet{ZPP68} \\
NGC 6334I, Sgr B2         &\citet{TWB70} \\
W3 IRS5, ON 1         &\citet{B74}   \\
Mon R2 IRS 3, W51, W58a/K3-50, DR21(OH), G111.542+0.777         &\citet{GMP83} \\
IRAS 06055+2039, IRAS 21413+5442, IRAS 12073--6233, G353.410--0.360        &\citet{CMW91} \\
DR21(OH)N         &\citet{CMC95} \\
G338.925+0.557    &\citet{M97}   \\
G294.511--1.621   &\citet{S97}   \\             
W75N, IRAS 22543+6145      &\citet{SKH00} \\
G240.316+0.071, G240.311+0.074, G328.307+0.430    &\citet{DE02}  \\
G328.808+0.633, G333.135--0.431, G011.904--0.141   &\citet{DE02}  \\
G338.075+0.012, G347.628+0.149    &\citet{S03}   \\
G043.148+0.013    &\citet{HSC05} \\
G034.84--0.95     &\citet{SXC18} \\
G173.482+2.446, G070.329+1.589, G006.049--1.447, G032.742--0.076   &\citet{QSB22} \\
G084.951--0.691   &\citet{OCS22} \\
G188.946+0.886, G196.454--1.677   &This paper  \\
\textbf{4660}  \\
Sgr B2            &\citet{TWB70}  \\
Orion IR          &\citet{RZP75}  \\
G351.775--0.536   &\citet{CMC95}  \\
G043.148+0.013    &\citet{PGW04}  \\
NGC 6334I         &\citet{MSG18}  \\
G012.209--0.102, G031.408+0.306, G031.213--0.180, W51               &\citet{QSB22}  \\
\textbf{4751} \\
I 06055+2039      &\citet{CMW91} \\
G348.550--0.979   &\citet{DE02}  \\   
\enddata
\end{deluxetable*}

\newpage
\section{4.7 GHz Gaussian parameters} \label{ApA}

\startlongtable
\begin{deluxetable*}{clccDll}
\tablecaption{Fitted Gaussian parameters to the 4.7\,GHz exOH Stokes $I$   
spectra. Columns contain (1) the ID number from Table \ref{tab:source}, (2) 
the source name, (3) the observing frequency, (4) the rms noise in the spectrum, 
(5) the peak flux density, (6) the velocity at the peak, and (7) the FWHM of 
the profile.}  
\label{tab:GaussFit47}  
\tablehead{ 
\colhead{No.}  &\colhead{Source}   &\colhead{Frequency}   &\colhead{RMS}    
         &\multicolumn2c{Peak Flux}   &\colhead{Velocity}   &\colhead{Width} \\
\colhead{}     &\colhead{Name}     &\colhead{(GHz)}       &\colhead{(mJy)}   
         &\multicolumn2c{Density (Jy)} &\colhead{(\kms)}    &\colhead{(\kms)} }
\decimalcolnumbers
\startdata
2a  &G111.526+0.803   &4.766 &8    &0.473(5)   &--60.272(2)  &0.494(6) \\
    &                 &      &     &0.114(4)   &--59.251(10) &0.53(3)  \\
    &                 &      &     &0.170(5)   &--58.328(8)  &0.52(2)  \\
    &                 &      &     &0.075(11)  &--57.50(7)   &0.50(12) \\
    &                 &      &     &0.182(15)  &--57.04(3)   &0.46(4)  \\
\hline    
2b  &G111.526+0.803   &4.766 &24   &0.386(10)  &--60.283(6)  &0.488(15) \\
    &                 &      &     &0.094(12)  &--59.26(2)   &0.39(6)   \\
    &                 &      &     &0.094(10)  &--58.36(3)   &0.58(8)   \\
    &                 &      &     &0.130(9)   &--57.22(2)   &0.73(6)   \\
\hline
3   &G111.532+0.759   &4.766 &17   &0.876(9)   &--60.269(3)  &0.473(6) \\
    &                 &      &     &0.170(8)   &--59.283(17) &0.64(5)  \\
    &                 &      &     &0.10(3)    &--58.510(15) &0.18(5)  \\
    &                 &      &     &0.165(11)  &--58.24(4)   &0.62(7)  \\
    &                 &      &     &0.272(8)   &--57.146(11) &0.72(3)  \\
\hline
4   &G111.542+0.777   &4.751 &12   &0.033(3)   &--59.06(18)  &4.3(4) \\
\\
    &                 &4.766 &16   &1.371(11)  &--60.272(2)  &0.480(4) \\
    &                 &      &     &0.288(11)  &--59.276(9)  &0.45(2)  \\
    &                 &      &     &0.232(13)  &--58.34(2)   &0.74(5)  \\
    &                 &      &     &0.13(2)    &--58.144(11) &0.15(3)  \\
    &                 &      &     &0.423(9)   &--57.190(8)  &0.67(2)  \\
\hline
6   &G133.715+1.215   &4.751 &23   &-0.054(8)  &+36.7(2)      &5.1(7) \\
    &                 &      &     &0.031(3)   &+45.2(12)     &12(3)  \\
\hline
6   &G133.715+1.215   &4.766 &27   &0.045(7)   &--39.81(14)  &1.9(3)  \\ 
    &                 &      &     &0.40(2)    &--37.854(5)  &0.220(13) \\
    &                 &      &     &0.095(18)  &--36.73(3)   &0.27(8) \\
    &                 &      &     &0.205(16)  &--36.306(17) &0.34(4) \\     
\hline 
7   &G133.947+1.064   &4.660 &15   &-0.042(4)  &--44.93(7)   &1.56(17) \\
\\
    &                 &4.751 &13   &0.023(9)   &--47.6(11)   &3.1(17) \\
    &                 &      &     &0.052(13)  &--44.9(4)    &2.7(5)  \\
\\
    &                 &4.766 &13   &0.025(4)   &--47.5(3)    &2.6(7)  \\
    &                 &      &     &4.492(13)  &--45.070(0)  &0.25(0) \\
    &                 &      &     &0.468(6)   &--44.641(17) &1.92(4) \\
    &                 &      &     &0.658(13)  &--43.542(4)  &0.30(1) \\
    &                 &      &     &0.581(13)  &--43.188(4)  &0.27(1) \\
\hline     
16a &Mon R2           &4.766 &18$^a$   &16.27(3)   &+10.7017(2)  &0.2854(5) \\
\hline
16b &Mon R2           &4.766 &26$^a$   &20.72(3)   &+10.6973(2)  &0.2797(5) \\
\hline
18  &G188.946+0.886   &4.766 &7   &0.237(4)   &+9.4872(14)   &0.197(3) \\
    &                 &      &    &0.119(3)   &+11.140(3)    &0.254(8) \\ 
\hline
19  &G196.454--1.677  &4.766 &17  &0.182(11)  &+18.816(9)   &0.25(2) \\
    &                 &      &    &0.059(10)  &+19.21(3)    &0.29(7) \\
    &                 &      &    &0.057(15)  &+19.658(17)  &0.14(4) \\
    \hline
$^a$   &\multicolumn7l{The rms values are different to \citet{SF25}, because 
\citetalias{SF25} } \\
      &\multicolumn7l{rms values should have been scaled down by a factor 2.17. In addition, rms } \\
      &\multicolumn7l{values presented here are reduced further due to Hanning smoothing.}
\enddata
\end{deluxetable*}

\section{6.0 GHz Gaussian parameters} \label{ApB}
\startlongtable
\begin{deluxetable*}{clccccDll} 
\tablecaption{Fitted Gaussian parameters to 6.0\,GHz exOH Stokes $I$, RCP and LCP 
spectra. Columns contain (1) the ID number from Table \ref{tab:source}, (2) the 
source name, (3) the observing frequency, (4) the polarization component 
of the fitted Gaussian, (5) the rms noise in the spectrum, (6) the number of the 
fitted Gaussian (7) the peak flux density, (8) the velocity at the peak, and (9) 
the FWHM of the profile. }
\label{tab:GaussFit6} 
\tablehead{ 
\colhead{No.}  &\colhead{Source}   &\colhead{Frequency}   &\colhead{Polarization}   
     &\colhead{RMS}    &\colhead{Component} &\multicolumn2c{Peak Flux}    &\colhead{Velocity}   
     &\colhead{Width} \\
\colhead{}     &\colhead{Name}     &\colhead{(GHz)}       &\colhead{Component}   
     &\colhead{(mJy)}  &\colhead{number} &\multicolumn2c{Density (Jy)} &\colhead{(\kms)}    
     &\colhead{(\kms)} }
\decimalcolnumbers
\startdata
1   &G108.758--0.986  &6.035 &RCP  &18  &1    &0.925(10)   &--45.939(2)    &0.293(5) \\
    &                 &      &     &    &2    &0.797(9)    &--45.537(2)    &0.306(6) \\
    &                 &      &LCP  &16  &1    &0.371(7)    &--45.405(3)    &0.340(9) \\
    &                 &      &     &    &2    &0.366(8)    &--44.925(3)    &0.278(8) \\
\hline
2a  &G111.526+0.803   &6.035 &I   &15   &1    &0.617(10)   &--59.497(2)    &0.219(5) \\
    &                 &      &    &     &2    &0.264(8)    &--59.123(7)    &0.352(18) \\
    &                 &      &    &     &3    &0.065(10)   &--58.570(16)   &0.21(4)  \\
2b  &G111.526+0.803   &6.035 &I   &28   &1    &0.42(3)     &--59.514(5)    &0.178(13)  \\
    &                 &      &    &     &2    &0.162(14)   &--59.15(3)     &0.44(7)  \\
\hline
3   &G111.532+0.759   &6.031 &I   &19   &1    &0.049(7)    &--59.18(6)     &0.93(15) \\
\\
    &                 &6.035 &I   &19   &1    &1.23(2)     &--59.505(2)    &0.202(4) \\
    &                 &      &    &     &2    &0.480(14)   &--59.145(7)    &0.41(2)  \\
    &                 &      &    &     &3    &0.053(11)   &--59.09(10)    &2.0(3)  \\
\\
    &                 &      &RCP &16   &1    &0.59(2)     &--59.497(3)    &0.182(7) \\
    &                 &      &    &     &2    &0.251(10)   &--59.190(17)   &0.44(4) \\
    &                 &      &    &     &3    &0.029(6)    &--59.01(16)    &2.5(4)  \\
\\
    &                 &      &LCP &14   &1    &0.640(12)   &--59.513(2)     &0.215(5) \\
    &                 &      &    &     &2    &0.244(11)   &--59.110(8)     &0.40(3) \\
    &                 &      &    &     &3    &0.019(9)    &--59.1(2)       &1.8(6) \\
\hline
4   &G111.542+0.777   &6.031 &I   &20   &1    &0.064(5)    &--59.02(6)     &1.59(15) \\
    \\ 
    &                 &6.035 &I   &23   &1    &2.559(17)   &--59.516(1)    &0.203(2)  \\
    &                 &      &    &     &2    &0.909(16)   &--59.149(3)    &0.372(9)  \\
    &                 &      &    &     &3    &0.143(12)   &--59.12(4)     &1.73(10)  \\
\\
    &                 &      &RCP &18   &1    &1.221(15)   &--59.5052(14)   &0.199(3) \\
    &                 &      &    &     &2    &0.446(16)   &--59.173(5)     &0.336(16) \\
    &                 &      &    &     &3    &0.102(15)   &-59.15(4)       &1.28(11) \\
\\
    &                 &      &LCP &16   &1    &1.359(10)   &--59.5255(9)     &0.205(2) \\
    &                 &      &    &     &2    &0.462(9)    &--59.123(3)      &0.370(10) \\
    &                 &      &    &     &3    &0.063(6)    &--59.04(5)       &2.06(15) \\
\hline
6   &G133.715+1.215   &6.035 &I   &25   &1    &0.173(5)    &+32.35(5)     &3.71(12)  \\
\hline
6   &G133.715+1.215   &6.017 &I   &27   &1    &-0.082(5)   &--39.810(8)   &2.6(2) \\
\\
    &                 &6.031 &I   &27   &1    &-0.053(12)  &--40.7(4)     &7.9(9) \\
    &                 &      &    &     &2    &-0.320(12)  &--39.45(3)    &2.98(10) \\
\\
    &                 &6.035 &I   &28   &1    &-0.046(7)   &--41.4(5)     &10.1(9) \\
    &                 &      &    &     &2    &-0.467(8)   &--39.52(2)    &3.09(6) \\
\hline
7   &G133.947+1.064   &6.017 &I   &13   &1    &-0.020(6)   &--47.67(11)   &0.8(3)  \\
    &                 &      &    &     &2    &-0.056(4)   &--45.54(5)    &1.33(12)  \\
    \\
    &                 &6.031 &I   &32   &1    &0.58(2)     &--47.402(3)   &0.237(8)  \\
    &                 &      &    &     &2    &0.66(2)     &--46.882(3)   &0.222(7)  \\
    &                 &      &    &     &3    &1.01(1)     &--45.843(4)   &0.326(9)  \\
    &                 &      &    &     &4    &1.83(3)     &--45.394(5)   &0.339(13) \\
    &                 &      &    &     &5    &1.90(5)     &--45.045(5)   &0.323(12) \\
    &                 &      &    &     &6    &1.45(1)     &--44.557(5)   &0.546(16) \\
    &                 &      &    &     &7    &5.72(11)    &--43.521(3)   &0.243(4)  \\
    &                 &      &    &     &8    &12.18(16)   &--43.337(1)   &0.202(2)  \\
    &                 &      &    &     &9    &6.26(12)    &--43.321(3)   &1.052(10) \\
    &                 &      &    &     &10   &13.65(12)   &--42.956(1)   &0.239(2)  \\
    &                 &      &    &     &11   &24.73(10)   &--42.742(1)   &0.230(1)  \\ 
    &                 &      &    &     &12   &7.32(3)     &--42.182(1)   &0.239(2)  \\ 
    &                 &      &    &     &13   &4.07(2)     &--41.913(2)   &0.261(3)  \\ 
\\
    &                 &      &RCP &29   &1    &14.3(12)    &--42.95(3)    &0.33(3) \\
    &                 &      &    &     &2    &24.(3)      &--42.728(8)   &0.237(10) \\
\\
    &                 &      &LCP &27   &1    &11.7(3)     &--43.465(7)    &0.488(7) \\
    &                 &      &    &     &2    &9.6(5)      &--43.340(2)    &0.200(8) \\
\\
    &                 &6.035 &I   &30   &1    &2.29(13)    &--48.841(6)   &0.214(15) \\
    &                 &      &    &     &2    &5.5(2)      &--47.825(3)   &0.210(10) \\
    &                 &      &    &     &3    &1.8(4)      &--47.30(2)    &0.35(8) \\
    &                 &      &    &     &4    &2.2(4)      &--47.17(7)    &1.6(2) \\
    &                 &      &    &     &5    &3.8(2)      &--46.939(9)   &0.269(18) \\
    &                 &      &    &     &6    &2.1(2)      &--46.137(12)  &0.30(3) \\
    &                 &      &    &     &7    &6.87(15)    &--45.603(6)   &0.50(2) \\
    &                 &      &    &     &8    &22.46(12)   &--45.0667(16) &0.377(4) \\
    &                 &      &    &     &9    &16.84(20)   &--44.432(4)   &0.502(9) \\
    &                 &      &    &     &10   &33.(8)       &--43.51(9)    &0.81(7) \\
    &                 &      &    &     &11   &61.(4)       &--43.286(3)   &0.262(5) \\
    &                 &      &    &     &12   &62.(5)       &--42.937(3)   &0.276(7) \\
    &                 &      &    &     &13   &45.(7)       &--42.80(10)   &0.83(15) \\
    &                 &      &    &     &14   &35.0(16)    &--42.075(3)   &0.421(7) \\
\\
    &                 &      &RCP &24   &1    &62.4(11)    &--42.988(4)   &0.456(11) \\
    &                 &      &    &     &2    &42.2(10)    &--42.149(6)   &0.537(17) \\
\\
    &                 &      &LCP &21   &1    &70.0(15)    &--43.333(3)   &0.309(9) \\
    &                 &      &    &     &2    &44.3(7)     &--42.860(10)  &0.58(2) \\
\\
    &                 &6.049 &I   &16   &1    &0.116(10)   &--45.195(15)  &0.40(4)  \\
    &                 &      &    &     &2    &0.085(8)    &--45.112(45)  &1.94(13) \\
\hline
8   &G141.918+1.902   &6.031 &I   &18   &1    &0.061(14)   &--8.15(2)     &0.19(5) \\ 
\\
    &                 &6.035 &I   &18   &1    &0.283(12)   &--8.090(5)    &0.268(13) \\ 
\\    
    &                 &      &RCP &14   &1    &0.050(10)   &--7.85(2)     &0.22(5) \\ 
\\
    &                 &      &LCP &13   &1    &0.290(9)    &--8.104(4)    &0.238(8) \\
\hline
15  &G183.349--0.575  &6.035 &I   &18   &1    &0.083(11)   &--5.72(2)     &0.34(6)   \\
    &                 &      &    &     &2    &0.290(14)   &--5.292(5)    &0.200(14) \\
    &                 &      &    &     &3    &0.110(10)   &--4.87(2)     &0.40(5)   \\
\\
    &                 &      &RCP &14   &1    &0.074(8)    &--5.744(18)   &0.33(4) \\  
    &                 &      &    &     &2    &0.041(13)   &--5.31(2)     &0.13(5) \\ 
\\
    &                 &      &LCP &12   &1    &0.249(8)    &--5.286(4)    &0.236(11) \\
    &                 &      &    &     &2    &0.116(7)    &--4.879(10)   &0.30(3) \\
\hline
17  &G189.030+0.783   &6.035 &I   &13   &1    &0.592(7)    &+3.3660(17)   &0.285(4) \\
    &                 &      &    &     &2    &0.309(8)    &+8.793(3)     &0.249(7) \\ 
\\
    &                 &      &RCP &11   &1    &0.363(6)    &+3.354(2)     &0.265(5) \\
    &                 &      &    &     &2    &0.124(7)    &+8.854(6)     &0.218(13) \\
\\
    &                 &      &LCP &10   &1    &0.237(5)    &+3.386(3)     &0.307(8) \\
    &                 &      &    &     &2    &0.219(6)    &+8.758(3)     &0.230(7) \\
\hline
21a &G240.316+0.071   &6.035 &I   &19   &1    &0.753(12)   &+62.7634(17)  &0.213(4) \\
    &                 &      &    &     &2    &2.204(13)   &+63.6150(7)   &0.1774(15) \\
    &                 &      &    &     &3    &0.465(11)   &+63.919(4)    &0.274(10) \\
\\
    &                 &      &RCP &13   &1    &0.405(8)    &+62.757(2)    &0.216(5) \\
    &                 &      &    &     &2    &0.489(10)   &+63.612(2)    &0.171(5) \\
    &                 &      &    &     &3    &0.234(8)    &+63.900(5)    &0.249(13) \\
\\
    &                 &      &LCP &12   &1    &0.349(7)    &+62.770(2)    &0.208(5) \\
    &                 &      &    &     &2    &1.719(8)    &+63.6165(5)   &0.1801(12) \\
    &                 &      &    &     &3    &0.243(7)    &+63.942(4)    &0.274(11) \\
\\
21b &G240.316+0.071   &      &I   &17   &1    &0.737(11)   &+62.7552(15)  &0.208(4) \\
    &                 &      &    &     &2    &2.211(12)   &+63.6124(6)   &0.1778(14) \\
    &                 &      &    &     &3    &0.454(10)   &+63.915(4)    &0.271(9) \\
\\
    &                 &      &RCP &13   &1    &0.398(8)    &+62.752(2)    &0.211(5) \\
    &                 &      &    &     &2    &0.511(9)    &+63.6161(19)  &0.180(4) \\
    &                 &      &    &     &3    &0.226(8)    &+63.910(5)    &0.237(12) \\
\\
    &                 &      &LCP &12   &1    &0.340(7)    &+62.759(2)     &0.206(5) \\
    &                 &      &    &     &2    &1.699(8)    &+63.6115(5)    &0.1775(12) \\
    &                 &      &    &     &3    &0.233(6)    &+63.921(5)     &0.300(13) \\
\enddata
\end{deluxetable*}

\section{Fitted Gaussian parameters for Mon R2} \label{ApC}
\begin{table*}[ht]
	\centering
	\caption{4.7 and 6.0\,GHz parameters of the thermal absorption and emission 
    lines observed in Mon R2}
	\label{tab:MonR2}
	\begin{tabular}{ccccDDD} 
		\hline
Frequency  &Number of  &Total Time   &RMS    &\multicolumn2c{Flux density}   
     &\multicolumn2c{Velocity}   &\multicolumn2c{FWHM}    \\    
(GHz)     &observations    &(mins)   &(mJy)    &\multicolumn2c{(Jy)}     
    &\multicolumn2c{\kms}   &\multicolumn2c{\kms}  \\
		\hline
4.660    &15   &864    &4   &-0.0077(10)  &10.92(13)       &2.1(3) \\
4.751    &18   &1024   &4   &0.0248(10)   &11.14(4)        &2.02(9) \\
4.766    &18   &1024   &5   &0.0876(12)$^b$   &11.139(11)$^b$  &1.71(3) \\
\hline
6.017    &10   &518    &6   &-0.0197(13)  &10.58(8)        &2.55(19) \\
6.031    &18   &1040   &5   &-0.046(2)    &9.14(18)        &4.5(2)   \\
         &     &       &    &-0.068(5)    &11.16(2)        &2.08(10) \\
6.035    &18   &1040   &5   &-0.025(5)    &8.31(5)         &1.3(2)   \\  
         &     &       &    &-0.048(7)    &9.18(15)        &5.1(2)  \\
         &     &       &    &-0.0774(6)   &11.04(4)        &2.09(11) \\ 
\hline
$^b$   &\multicolumn9l{Values slightly different to \citetalias{SF25} due to refinement 
in weighting of the } \\
   &\multicolumn9l{observations.}  \\ 
\hline
	\end{tabular}
\end{table*}

\end{document}